\documentclass[journal]{IEEEtran}

\usepackage{graphicx,amssymb,lineno}
\usepackage{amsmath,amsfonts,amssymb}

\usepackage{algorithm}
\usepackage{algorithmic}
\usepackage[usenames]{color}
\usepackage{float}

\usepackage{graphicx,graphics,color,epsfig,subfigure,graphpap,rotate}
\usepackage{times, verbatim, subfigure, epsfig, graphicx, latexsym, amsmath}
\usepackage{url}
\usepackage{subfigure}

\newtheorem{remark}{Remark}

\begin{document}
%
\title{Exploiting Device-to-Device Communications in Joint Scheduling of Access and Backhaul for mmWave Small Cells}

\author{Yong~Niu,
         Chuhan~Gao,
         Yong~Li,~\IEEEmembership{Member,~IEEE,}
         Li~Su,
         Depeng~Jin,~\IEEEmembership{Member,~IEEE,}
        and Athanasios V. Vasilakos,~\IEEEmembership{Senior Member,~IEEE}
\thanks{Y. Niu, C. Gao, Y. Li, D.~Jin and L.~Su are with State Key Laboratory on
 Microwave and Digital Communications, Tsinghua National Laboratory for Information
 Science and Technology (TNLIST), Department of Electronic Engineering, Tsinghua
 University, Beijing 100084, China (E-mails: liyong07@tsinghua.edu.cn).} 
\thanks{A. V. Vasilakos is with the Department of Computer and Telecommunications Engineering,
 University of Western Macedonia, Greece.} %
\thanks{This work was partially supported by the National Natural Science
Foundation of China (NSFC) under grant No. 61201189 and 61132002, National High Tech (863) Projects
under Grant No. 2011AA010202, Research Fund of Tsinghua University under No. 2011Z05117 and
20121087985, and Shenzhen Strategic Emerging Industry Development Special Funds under No.
CXZZ20120616141708264.}
}%

\maketitle

\begin{abstract}

With the explosive growth of mobile data demand, there has been an increasing interest in deploying
small cells of higher frequency bands underlying the conventional homogeneous macrocell network,
which is usually referred to as heterogeneous cellular networks, to significantly boost the overall
network capacity. With vast amounts of spectrum available in the millimeter wave (mmWave) band, small cells
at mmWave frequencies are able to provide multi-gigabit access data rates, while the wireless backhaul in the mmWave band is emerging as a cost-effective solution to provide high backhaul capacity to
connect access points (APs) of the small cells. In order to operate the mobile network optimally,
it is necessary to jointly design the radio access and backhaul networks. Meanwhile, direct
transmissions between devices should also be considered to improve system performance and enhance
the user experience. In this paper, we propose a joint transmission scheduling scheme for the radio
access and backhaul of small cells in the mmWave band, termed D2DMAC, where a path selection criterion
is designed to enable device-to-device transmissions for performance improvement. In D2DMAC, a
concurrent transmission scheduling algorithm is proposed to fully exploit spatial reuse in mmWave
networks. Through extensive simulations under various traffic patterns and user deployments, we demonstrate D2DMAC achieves near-optimal performance in some cases,
and outperforms other protocols significantly in terms of delay and throughput. Furthermore, we
also analyze the impact of path selection on the performance improvement of D2DMAC under different
selected parameters.



\begin{IEEEkeywords}
Heterogeneous cellular networks, small cells, MAC scheduling, millimeter wave communications, 60
GHz, device-to-device.

\end{IEEEkeywords}

\end{abstract}

\section{Introduction}\label{S1}


Mobile traffic demand is increasing rapidly, and a 1000-fold demand increase by 2020 is predicted
by some industry and academic experts \cite{data1, data2}. In order to meet this explosive growth
demand and enhance the mobile network capacity, there has been an increasing interest in deploying
small cells underlying the conventional homogeneous macrocell network \cite{60GHz-backhaul-2}. This
new network deployment is usually referred to as heterogeneous cellular networks (HCNs). However,
reducing the radii of small cells in the carrier frequencies employed in today's cellular systems
to reap the spatial reuse benefits is fundamentally limited by interference constraints
\cite{Pico_60GHz}. By using higher frequency bands, such as the millimeter wave (mmWave) bands
between 30 and 300 GHz, and bringing the network closer to users by a dense deployment of small
cells \cite{Pico_60GHz,dense cells}, HCNs can significantly boost the overall network capacity due
to less interference and higher data rates. With huge bandwidth available in the mmWave band, small cells at mmWave frequencies are able to provide multi-gigabit communication
services, such as uncompressed and high-definition video transmission, high speed Internet access,
and wireless gigabit Ethernet for laptops and desktops, and have attracted considerable interest
from academia, industry, and standards bodies \cite{60GHz-1,60GHz-2}. Rapid development in
complementary metal-oxide-semiconductor radio frequency integrated circuits \cite{CMOS,CMOS2,CMOS3}
accelerates the industrialization of mmWave communications. Meanwhile, standards,
such as ECMA-387 \cite{ECMA 387} , IEEE 802.15.3c \cite{IEEE 802.15.3c}, and IEEE 802.11ad
\cite{IEEE 802.11ad}, have been defined for indoor wireless personal area networks (WPAN) or
wireless local area networks (WLAN).


Unlike existing communication systems of macrocells using lower carrier frequencies (e.g., from 900
MHz to 5 GHz), small cells in the mmWave band suffer from high propagation loss. The free space
propagation loss at 60 GHz band is 28 decibels (dB) more than that at 2.4 GHz \cite{singh_outdoor}.
To combat severe channel attenuation, high gain directional antennas are utilized at both the
transmitter and receiver to achieve directional transmission \cite{beam_training, Beamtraining2, MRDMAC}. With a small wavelength, it is feasible to produce low-cost and compact on-chip and in-package antennas in the mmWave band \cite{CMOS3, mao}. Under directional transmission, the third party nodes cannot perform carrier sense as in IEEE 802.11 to avoid contention with current transmission, which is referred to as the deafness problem. On the other hand, there is less interference among links, and concurrent transmissions (spatial reuse)
can be exploited to greatly improve network capacity. Due to directivity and high propagation loss,
the interference among cells is minimal, and total capacity of small cells scales with the number
of cells in a given region \cite{Pico_60GHz}. On the other hand, as the data transmission rates of
radio access networks increase dramatically, providing backhaul to transport data to (from) the
gateway in the core network for small cells becomes a significant challenge
problem \cite{60GHz-backhaul-2}. Although fiber based backhaul offers larger bandwidth to meet this
requirement, it is costly, inflexible and time-consuming to connect the densely deployed small
cells with high connectivity. In contrast, high speed wireless backhaul is more cost-effective,
flexible, and easier to deploy \cite{60GHz-backhaul-2, Narlikar, 60GHz-backhaul-1,
60GHz-backhaul-4}. Wireless backhaul in mmWave bands, especially 60 GHz band, can be a promising
backhaul solution for small cells \cite{60GHz-backhaul-1}. The wide bandwidth enables several-Gbps
data rate even with simple modulation schemes such as OOK, BPSK and FSK.



\begin{figure} [htbp] 
\begin{center}
\includegraphics*[width=9cm]{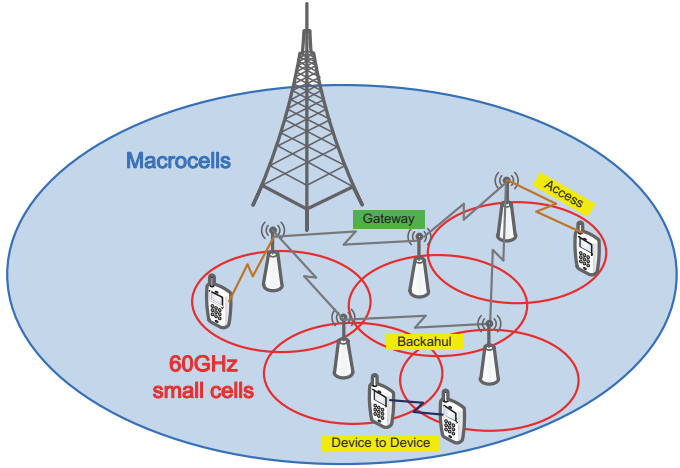}
\end{center}
\caption{Dense deployment of small cells in the 60 GHz band underlying the macrocell network.}
\label{small cell}
\end{figure}


In Fig. \ref{small cell}, we present a typical scenario for dense deployment of small cells in the
60 GHz band underlying the macrocell cellular network, which forms HCNs. In HCNs, the macrocell is coupled with the small cells in the mmWave band to some extent \cite{hybrid 4G}, and the macrocell network is allowed to have certain visibility and radio resource management control over the small cells \cite{HCN_book}. Users can communicate via the current macrocell as well as the small cells. In the small cells, mobile
users are associated with access points (APs), and APs are connected through wireless backhaul
links. Some APs are connected to the Internet via a direct and high speed wired connection, which
are called \emph{gateways}. In this targeted small cells system, both the radio access and backhaul
links are in the 60 GHz band, which can provide high data rate for both radio access and backhaul,
and also reduce implementation complexity and deployment cost. Considering the fundamental
differences between mmWave communications and existing other communication systems using lower
carrier frequencies, scheduling over both the radio access and backhaul networks is a difficult
problem and should be carefully designed. There are two aspects of challenges. In the first aspect,
it is necessary to consider both radio access and backhaul networks jointly \cite{joint_design}. On
one hand, co-design of scheduling for transmission over radio access and backhaul networks can
maximize the spatial reuse across radio access and backhaul networks. Consequently, network
capacity and performance can be improved significantly. On the other hand, joint scheduling for
transmission over radio access and backhaul networks can also manage the interference both within
each cell and among cells efficiently. In the second aspect, under dense deployment, there is a
high probability that two devices within both the same cell and different cells are located near to
each other. In this case, user devices can transmit data to each other over direct links using the
small cell resources, instead of through APs. Thus, most of context-aware applications that involve
discovering and communicating with nearby devices, including the popular content downloading, can
benefit from the device-to-device communications by reducing the communication cost since it
enables physical-proximity communication, which saves power while improving the spectral
efficiency \cite{hua1}. Therefore, the potential of device-to-device communications to enhance the network
performance should also be considered in the joint scheduling problem over both radio access and
backhaul networks.

Aiming to address the above challenges, in this paper, we propose a joint transmission scheduling
scheme, termed D2DMAC, for small cells in the mmWave band. In D2DMAC, joint scheduling over radio
access and backhaul networks is considered, and direct transmissions between devices are further
exploited to improve network performance in terms of throughput and delay. If the direct link
between the sender and receiver of one flow has high channel quality, the direct transmission will
be adopted instead of transmission through the backhaul network. The contributions of this paper
are three-fold, which are summarized as follows.


\begin{itemize}
\item We formulate the joint scheduling problem over both the radio access and backhaul networks with the direct transmissions between devices considered into a mixed integer
linear program (MILP), i.e., to minimize the number of time slots to accommodate the traffic
demand of all flows. Concurrent transmissions, i.e., spatial reuse, are explicitly considered in
this formulated problem.
\item We propose an efficient scheduling scheme termed D2DMAC, which consists of a path selection criterion and a transmission scheduling algorithm to solve the formulated problem.
The priority of device-to-device transmission is characterized by the path selection parameter of
the path selection criterion, while concurrent transmissions are fully exploited in the transmission
scheduling algorithm to maximize the spatial reuse gain.
\item We analyze the concurrent transmission conditions in D2DMAC, and derive a sufficient condition for each link to support concurrent transmissions with other links, i.e. , the distances between the interference sources and the receiver of this link should be larger than or equal to its interference radius. We also analyze the interference radii under different modulation and coding schemes, different path loss exponents, different link lengths, and different transmission power.
\item Through extensive simulations under different traffic patterns and user deployments, we demonstrate our proposed D2DMAC achieves near-optimal performance in some cases in terms of delay and throughput by comparison with the optimal solution, and
it outperforms the other existing related protocols significantly. Furthermore, we analyze the
impact of path selection parameter on performance improvement.

\end{itemize}

The paper is organized in the following way: Section \ref{S2} presents related work on scheduling
in the mmWave band. Section \ref{S3} introduces the system model and illustrates our basic idea by
an example. Section \ref{S4} formulates the optimal joint scheduling problem of access and backhaul
with device-to-device transmissions considered as an MILP. Section \ref{S5} presents our proposed
scheme, D2DMAC, which includes the path selection criterion and transmission scheduling algorithm.
Section~\ref{S5-i} presents the conditions for concurrent transmissions and gives a performance
analysis. Section \ref{S6} shows the simulation results of D2DMAC under various traffic patterns
and different path selection parameters, and the comparison with the optimal solution and other
protocols. Finally, we conclude this paper in Section \ref{S7}.

\section{Related Work}\label{S2}


Recently, some related work on scheduling for small cells in the mmWave band has been
published. Since the standards of ECMA-387 \cite{ECMA 387} and IEEE 802.15.3c \cite{IEEE 802.15.3c}
adopt TDMA, some work is also based on TDMA \cite{mao_12, mao_13, Qiao_6, Qiao_15, Qiao, EX_Region, Qiao_7}. In two
protocols based on IEEE 802.15.3c \cite{Qiao_6, Qiao_15}, multiple links are scheduled to
communicate in the same slot if the multi-user interference (MUI) between them is below a specific
threshold. Qiao \emph{et al.} \cite{Qiao} proposed a concurrent transmission scheduling algorithm
to maximize the number of flows scheduled with the quality of service requirement of each flow
satisfied. Cai \emph{et al.} \cite{EX_Region} introduced the concept of exclusive region (ER) to
support concurrent transmissions, and derived the conditions that concurrent transmissions always
outperform TDMA. Qiao \emph{et al.} \cite{Qiao_7} proposed a multi-hop concurrent transmission
scheme (MHCT) to address the link outage problem (blockage) and combat large path loss to improve
flow throughput. For bursty traffic, TDMA based protocols face the problem that the medium access
time for a flow is often highly unpredictable, which will result in over-allocated medium access
time for some users while under-allocated medium access time for others. Besides, TDMA based
protocols may also have high control overhead for medium reservation.

There is some other work based on a central coordinator to coordinate the transmissions in WPANs
\cite{MRDMAC, mao, mao_11, chenqian}. Gong \emph{et al.} \cite{mao_11} proposed a directional CSMA/CA
protocol, which exploits virtual carrier sensing, and depends on the piconet coordinator (PNC) to
distribute the network allocation vector (NAV) information. It mainly focuses on solving the
deafness problem and does not exploit the spatial reuse fully. Recently, Son \emph{et al.}
\cite{mao} proposed a frame based directional MAC protocol (FDMAC).
The high efficiency of FDMAC is achieved by amortizing the
scheduling overhead over multiple concurrent transmissions in a row.
The core of FDMAC is the Greedy Coloring algorithm, which fully
exploits spatial reuse and greatly improves the network throughput
compared with MRDMAC \cite{MRDMAC} and memory-guided directional MAC
(MDMAC) \cite{MDMAC}. FDMAC also has a good fairness performance
and low complexity. Singh \emph{et al.} \cite{MRDMAC}
proposed a multihop relay directional MAC protocol (MRDMAC), which overcomes the deafness problem
by PNC's weighted round robin scheduling. In MRDMAC, if a wireless terminal (WT) is lost due to
blockage, the access point (AP) will choose a live WT to act as a relay to the lost WT. Since most
transmissions are through the PNC, MRDMAC does not consider the spatial reuse. Chen \emph{et al.}
\cite{chenqian} proposed a directional cooperative MAC protocol, D-CoopMAC, to coordinate the
uplink channel access in an IEEE 802.11ad WLAN. D-CoopMAC selects a relay for a direct link, and
when the two-hop link outperforms the direct link, the latter will be replaced by the former.
Spatial reuse is also not considered in D-CoopMAC since most transmissions go through the AP. All
the work above considers scheduling in the access network, and does not consider the backhaul
network. For the small cell networks in the mmWave band, since the backhaul network becomes the
bottleneck, the backhaul and access networks should be considered jointly to promote the spatial
reuse and perform efficient interference management both within and among cells.


On the other hand, there is also some related work on wireless backhaul networks. Lebedev \emph{et
al.} \cite{60GHz-backhaul-1} identified advantages of the 60 GHz band for short-range mobile
backhaul through feasibility analysis and comparison with the E-band technology. Islam \emph{et
al.} \cite{60GHz-backhaul-2} performed the joint cost optimal aggregator node placement, power
allocation, channel scheduling and routing to optimize the wireless backhaul network in mmWave
bands. Bojic \emph{et al.} \cite{60GHz-backhaul-4} discussed advanced wireless and optical
technologies for small-cell mobile backhaul with dynamic software-defined management in 60 GHz band
and E-band. Bernardos \emph{et al.} \cite{joint_design} discussed the main challenges to design the
backhaul and radio access network jointly in a cloud-based mobile network. Design ideas involving
the physical layer, the MAC layer, and the network layer are proposed. Most of the work on the
backhaul above does not consider the access network, and fails to give a joint design of scheduling
over the access and backhaul networks. Besides, direct transmissions between devices are also not
exploited to improve network performance in the work above. To the best of our knowledge, we are
the first to consider the transmission scheduling over both access and backhaul networks jointly,
and also exploit the device-to-device transmission for performance improvement.

As regards small cells in the mmWave band, most work focused on using bands such as 28GHz, 38GHz
and 73GHz to attain ranges of the order of 200m or even more \cite{mmW-cellular,Sarnoff}. Zhu
\emph{et al.} \cite{Pico_60GHz} proposed a 60GHz picocell architecture to augment existing LTE
networks for significant increase in capacity. The feasibility of this architecture is investigated
by characterizing range, attenuation due to reflections, sensitivity to movement and blockage, and
interference in typical urban environments. As in Ref. \cite{Pico_60GHz}, the oxygen absorption
that peaks at 60GHz is not a barrier to high-capacity 60GHz outdoor picocells, and can even promote
spatial reuse due to less interference from faraway base stations.

\section{System Overview}\label{S3}

\subsection{System Model}\label{S3-1}

We consider the scenario that multiple small cells are deployed. In each small cell, there are
several wireless nodes (WNs) and an access point (AP), which synchronizes the clocks of WNs and
provides access services within the cell. The APs form a wireless backhaul network in the mmWave band,
which should be stationary. Also, the backhaul links between APs are able to be optimized in order
to achieve high channel quality and reduce interference. There are some APs connected to the
Internet via a direct and high speed wired connection, which are called \emph{gateways}. The
remaining APs should communicate with the gateways to send (receive) data to (from) Internet. The
backhaul network can form arbitrary topology like ring topology, star topology, or tree topology
\cite{ICTON 2012}, where the backhaul paths between APs should be optimized to maximize backhaul
efficiency. To overcome huge attenuation in the mmWave band, both the WNs and APs achieve directional transmissions with electronically steerable
directional antennas by the beamforming techniques \cite{MRDMAC, hua2}. In such a system of small cells networks, we
schedule the transmission of the access and backhaul networks jointly, which apparently differs
from the scheduling problem of the ad-hoc network.


In this paper, we adopt centralized control. On one hand, distributed control does not scale well \cite{SoftRAN}, and the latency will increase significantly with the number of APs, which is unsuitable for mmWave communication systems where a time slot only lasts for a few microseconds. Moreover, distributed control is difficult to achieve intelligent control mechanisms required for the complex operational environments that involve dynamic behaviors of accessing users and temporal variations of the communication links. On the other hand, centralized control is able to achieve flexible control in an optimized manner \cite{mobicom_poster}. There is a centralized controller in the network, which usually resides on a gateway and is similar to the central coordinator in IEEE 802.11ad \cite{IEEE 802.11ad}. The system is partitioned into non-overlapping time slots of equal length, and the controller
synchronizes the clocks of APs. Then the clocks of WNs are synchronized by their corresponding APs.

There is a bootstrapping program in the system, by which the central controller knows the up-to-date network topology and the location information of APs and WNs. The network topology can be obtained by the neighbor discovery schemes in \cite{bootstrapping,neighbor discovery,neighbor discovery 2,neighbor discovery 3,neighbor discovery 4}. Location information of nodes can be obtained based on wireless channel signatures, such as angle of arrival (AoA), time difference of arrival (TDoA), or the received signal strength (RSS) \cite{location_1, location_2, location_3, location_4, location_5}. In our system, the bootstrapping program adopts the direct discovery scheme to discover the network topology \cite{bootstrapping}. In the direct discovery scheme, a node is in the transmitting or receiving state at the beginning of each time slot. In the transmitting state, a node transmits a broadcast packet with its identity in a randomly chosen direction. In the receiving state, a node listens for broadcast packets from a randomly chosen direction. If a collision happens, the node fails to discover any neighbor; otherwise, if the transmitter is unknown, the receiver discovers a new neighbor by recording the angle of arrival (AOA) and the transmitter's identity. After the direct discovery, the nodes report discovered neighbors to the APs, and then APs report obtained topology information to the central controller. At the same time, the APs obtain the location information of WNs by a maximum-likelihood (ML) classifier based on changes in the second-order statistics and sparsity patterns of the beamspace multiple input multiple output (MIMO) channel matrix \cite{location_1}. Besides, in the heterogeneous cellular networks scenario, the location information can also be obtained by the cellular geolocation from cellular base stations in \cite{location_2} if the macrocell is coupled with the small cells to some extent. Then the central controller obtains the location information of nodes from the APs. With relatively low mobility, the network topology and location information will be updated periodically.

There are two kinds of flows transmitted in the network, flows between WNs
and flows from or to the Internet (gateway). Specifically, we assume there are $N$ flows with
traffic transmission demand in the network. For flow $i$, its traffic demand is denoted as $d_i$.
The traffic demand vector for all flows is denoted as $\textbf{d}$, a $1\times N$ matrix whose
$i$th element is $d_i$. For each flow, there are two possible transmission paths in our system, the
ordinary path and direct path. The ordinary path is the transmission path through APs, which may
includes the access link from the source to its associated AP, the backhaul path from the source's
associated AP to the destination's associated AP or the gateway, and the access link from the
destination's associated AP to the destination. We also assume the backhaul path from the source's
associated AP to the destination's associated AP across the backhaul network is pre-determined by
some criterion, such as the criterion of minimum number of hops, or maximum minimum transmission
rate on the backhaul path. For the $j$th hop link of the ordinary path for flow $i$, we denote its
sender as $s_{ij}$ and receiver as $r_{ij}$, and denote this link as $(s_{ij},r_{ij})$. The direct
path is the direct transmission path from the source to the destination, which does not pass
through the backhaul network and only has one hop. We denote the direct link of flow $i$ as
$(s_{i}^d,r_{i}^d)$, where $s_{i}^d$ is the source, and $r_{i}^d$ is the destination.

 For each flow $i$, its transmission rate vector on the ordinary path
is denoted as ${\bf{c}}_i^b$, where each element $c_{ij}^b$ represents the transmission rate of the
$j$th hop on the ordinary path. The maximum number of hops of the ordinary paths is denoted by
$H_{max}$. We also denote the transmission rate matrix for the ordinary paths of flows as
$\textbf{C}^b$, where the $i$th row of $\textbf{C}^b$ is ${\bf{c}}_i^b$ and $\textbf{C}^b$ is an
$N\times H_{max}$ matrix. The transmission rate of the direct path for flow $i$ is denoted as
$c_{i}^d$, and the transmission rate vector for the direct paths of flows is denoted as
$\textbf{c}^d$, a $1\times N$ matrix whose $i$th element is $c_{i}^d$.


In the LOS case, the achievable maximum
transmission rates of links can be estimated according to Shannon's channel capacity \cite{Qiao}.
In our system, the transmission rates on the ordinary paths and direct paths can be obtained via a
channel transmission rate measurement procedure \cite{tvt_own}. In this procedure, the transmitter
of each link transmits measurement packets to the receiver first. Then with the measured signal to
noise ratio (SNR) of received packets, the receiver estimates the achievable transmission rate, and
appropriate modulation and coding scheme by the correspondence table about SNR and modulation and
coding scheme. Under low user mobility, the procedure is usually executed periodically. Besides,
since the APs are not mobile and positioned high above to avoid blockage \cite{Pico_60GHz}, the
transmission rates of backhaul links can be assumed fixed for a relatively long period of time.




With directional transmissions, there is less interference between links. Under low multi-user
interference (MUI) \cite{Qiao}, concurrent transmissions can be supported. In the system, all nodes
are assumed to be half-duplex, and two adjacent links cannot be scheduled concurrently since each
node has at most one connection with one neighbor \cite{mao}. For two nonadjacent links, we adopt
the interference model in \cite{Qiao} and \cite{Xu_mis}. For link $(s_i,r_i)$ and link $(s_j,r_j)$,
the received power from $s_i$ to $r_j$ can be calculated as
\begin{equation}
P_{r_j,s_i} = {f_{s_i,r_j}}k_0{P_t}{l_{s_ir_j}}^{ - \gamma },
\end{equation}
where ${{P_t}}$ is the transmission power that is fixed, $k_0 = {10^{PL({d_0})/10}}$ is the constant
scaling factor corresponding to the reference path loss ${PL({d_0})}$ with $d_0$ equal to 1 m,
${{l_{s_ir_j}}}$ is the distance between node $s_i$ and node $r_j$, and $\gamma $ is the path loss
exponent \cite{Qiao}. $f_{s_i,r_j}$ indicates whether $s_i$ and $r_j$ direct their beams towards
each other. If it is, ${{f_{s_i,r_j}}}=1$; otherwise, ${{f_{s_i,r_j}}}=0$. In other words, ${{f_{s_i,r_j}}}=0$ if and only if any transmitter is outside
the beamwidth of the other receiver or does not
direct its beam to the other receiver if it is within
the beamwidth of the other receiver \cite{Qiao_D2D}. Thus, the SINR at $r_j$,
denoted by $SINR_{s_jr_j}$, can be calculated as
\begin{equation}
SIN{R_{s_jr_j}} = \frac{{k_0{P_t}{l_{s_jr_j}}^{ - \gamma }}}{{W{N_0} + \rho \sum\limits_{i \ne j}
{{f_{s_i,r_j}}k_0{P_t}{l_{s_ir_j}}^{ - \gamma }} } },
\end{equation}
where $\rho$ is the MUI factor related to the cross correlation of signals from different links,
$W$ is the bandwidth, and ${{N_0}}$ is the onesided
power spectra density of white Gaussian noise \cite{Qiao}. For link $(s_i,r_i)$, the minimum SINR to support its transmission rate
${c_{s_i,r_i}}$ is denoted as $MS({c_{s_ir_i}})$. Therefore, concurrent transmissions can be
supported if the SINR of each link $(s_i,r_i)$ is larger than or equal to $MS({c_{s_ir_i}})$.

\subsection{D2DMAC Operation Overview}
The operation of D2DMAC is illustrated in Fig. \ref{frame}. D2DMAC is a frame based MAC protocol as
FDMAC \cite{mao}. Each frame consists of a scheduling phase and a transmission phase, and the scheduling overhead in the scheduling phase can be amortized over multiple concurrent transmissions in the transmission phase as in \cite{mao}. In the
scheduling phase, after the WNs of each cell steer their antennas towards their associated AP, each
AP polls its associated WNs successively for their traffic demand. Then each AP reports the traffic
demand of its associated WNs to the central controller on the gateway through the backhaul network.
We denote the time for the central controller to collect the traffic demand as ${t_{poll}}$. Based
on the transmission rates of links, the central controller computes a schedule to accommodate the
traffic demand of all flows, which takes time ${t_{sch}}$. Then the central controller pushes the
schedule to the APs through the backhaul network. Afterwards, each AP pushes the schedule to its
associated WNs. We denote the time for the central controller to push the schedule to the APs and
WNs as ${t_{push}}$. In the transmission phase, WNs and APs communicate with each other following
the schedule until the traffic demand of all flows is accommodated. The transmission phase consists
of multiple stages, and in each stage, multiple links are activated simultaneously for concurrent
transmissions. In the schedule computation, first the transmission path should be selected
optimally between the direct path and ordinary path for each flow. Then, the schedule should
accommodate the traffic demand of flows with a minimum number of time slots, which indicates
concurrent transmissions (spatial reuse) should be fully exploited.

\begin{figure} [htbp] 
\begin{center}
\includegraphics*[width=8.5cm]{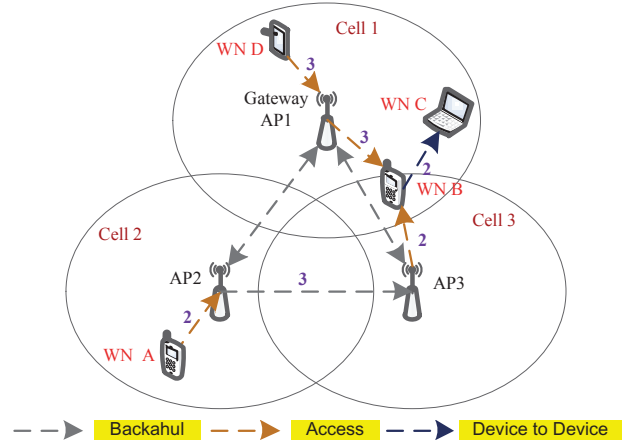}
\end{center}
\caption{An example of D2DMAC with three cells.} \label{scenario}
\end{figure}

\subsection{Problem Overview}\label{S3-2}

To maximize the transmission efficiency in the above introduced operation procedure, the traffic
demand of flows should be accommodated with a minimum number of time slots in the transmission
phase. To achieve this, first we should unleash the potential of spatial reuse to enable as many
link to transmit concurrently as possible. Meanwhile, for flows with a direct link of high quality,
we prefer the direct path to the ordinary path. In this case, we should enable the direct
transmission from the source to the destination to improve flow and network throughput.

\begin{figure} [htbp] 
\begin{center}
\includegraphics*[width=8.5cm]{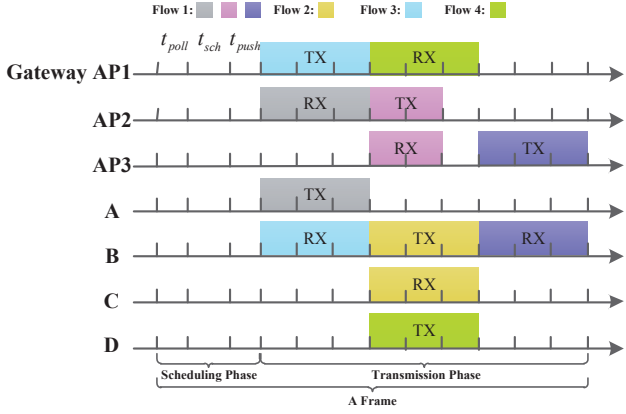}
\end{center}
\caption{Time-line illustration of D2DMAC operation.} \label{frame}
\end{figure}

Now, we present an example to illustrate the basic idea of D2DMAC. In Fig.~\ref{scenario}, there
are three small cells. In cell 1, node C and D are associated with AP1; in cell 2, node A is
associated with AP2; in cell 3, node B is associated with AP3. There are four flows in the network,
A $\to$ B, B $\to$ C, gateway $\to$ B, and D $\to$ gateway. The traffic demands of A $\to$ B, B
$\to$ C, gateway $\to$ B, and D $\to$ gateway are 5, 6, 7, and 8, respectively, and thus
$\textbf{d}=[5\;6\; 7 \;8]$. Numerically, they are equal to the number of packets to be
transmitted, and the packet length is fixed. After the channel transmission rate measurement
procedure, the obtained transmission rate matrix for the ordinary paths of flows is
\begin{equation}
\textbf{C}^b= \left(
\begin{aligned}
     2 &&~~ 3 &&~~ 2 \\
     2 &&~~ 4 &&~~ 2 \\
     4 &&~~ 2 &&~~ 0 \\
     0 &&~~ 0 &&~~ 0 \\
 \end{aligned}
 \right),\label{eqn:C_b}
\end{equation}
which indicates the transmission rates of link A $\to$ AP2, AP2 $\to$ AP3, and AP3 $\to$ B are 2,
3, and 2 respectively. Numerically, the transmission rates are equal to the number of packets these
links can transmit in one time slot. Since the flow from D to the gateway (AP1) does not need to go
through the backhaul network, the fourth row of $\textbf{C}^b$ is set to 0. The transmission rate
vector for the direct paths of flows is $\textbf{c}^d=[1\;2\; 3 \;3]$, and it indicates the direct
link of A $\to$ B can transmit one packet in one time slot; the direct link AP1 $\to$ B can
transmit three packets in one time slot. For each flow, first we should select optimal transmission
path between its direct path and ordinary path. Then the optimal schedule should accommodate the
traffic demand of flows with a minimum number of time slots. Thus concurrent transmission should be
fully exploited in the schedule. Due to the half-duplex assumption, adjacent links cannot be
scheduled concurrently. Due to the inherent order of transmissions, preceding hops on each path
should be scheduled ahead of hops behind. Besides, the SINR of each link scheduled for concurrent
transmissions in a stage should be able to support its transmission rate.

If we select flow A $\to$ B to be transmitted through its ordinary path, and three other flows to
be transmitted through their direct paths, we can obtain a schedule to accommodate the traffic
demand of four flows. The schedule consists of three stages, and is already illustrated in Fig.
\ref{frame}. In Fig.~\ref{scenario}, we plot the links that are activated in the schedule of Fig.
2, and their transmission rates are labeled above these links. In the first stage, access links AP1
$\to$ B and A $\to$ AP2 transmit for three time slots; in the second stage, backhaul link AP2 $\to$
AP3 transmits for two time slots, direct link B $\to$ C for three time slots, and access link D
$\to$ AP1 for three time slots; in the third stage, access link AP3 $\to$ B transmits for three
time slots. In each stage, the SINR of each link is able to support its transmission rate. This
schedule needs nine time slots to accommodate the traffic demand of four flows. Conversely, if the
traffic of flow A $\to$ B is transmitted through the direct path, this flow needs five time slots
to be accommodated. Since direct link B $\to$ C, access link AP1 $\to$ B, and direct link A $\to$ B
have common node B, i.e., are adjacent, they cannot transmit concurrently in one stage. In this
case, it needs at least eleven time slots to accommodate the traffic demand of these three flows.
Therefore, the selection of transmission paths for flows has a significant impact on the efficiency
of scheduling. Besides, concurrent transmission scheduling also needs to be optimized to improve
transmission efficiency.

\section{Problem Formulation And Analysis}\label{S4}

\subsection{Problem Formulation} \label{S4-1}

 We denote a schedule as $\textbf{S}$, and
assume it has $K$ stages. In each stage of a schedule, multiple links are scheduled to transmit
concurrently. For each flow $i$, we define a binary variable $a_i^k$ to indicate whether the direct
link of flow $i$ is scheduled to transmit in the $k$th stage. If it is, $a_i^k=1$; otherwise,
$a_i^k=0$. We also define a binary variable $b_{ij}^k$ to indicate whether the $j$th hop of the
ordinary path of flow $i$ is scheduled to transmit in the $k$th stage. In the problem, we only
consider flows that have at least one unblocked path, the direct path or the ordinary path. The
number of time slots of the $k$th stage is denoted as $\delta ^k$. For each flow $i$, we define the
number of hops of its ordinary path as its hop number $H_i$; if flow $i$ does not have the ordinary
path as flow D $\to$ gateway in Fig. \ref{scenario} , $H_i$ will be set to 1. For any two links
$(s_i,r_i)$ and $(s_j,r_j)$, we define a binary variable $I(s_i,r_i,s_j,r_j)$ to indicate whether
these two links are adjacent. If they are, $I(s_i,r_i,s_j,r_j)=1$; otherwise,
$I(s_i,r_i,s_j,r_j)=0$. In a schedule, if a link is scheduled in one stage, it will transmit as
many packets as possible until its traffic demand is cleared. Then, the link will not be active in
the remaining slots of this stage.

Given the traffic demand of flows, to maximize transmission efficiency, we should accommodate the
traffic demand with a minimum number of time slots \cite{mao}. The total number of time slots of a schedule is
$\sum\limits_{k = 1}^K {{\delta ^k}} $. Now, we analyze the system constraints for a schedule.
First, we define a binary variable $h_{ij}$ to indicate whether flow $i$ has traffic to transmit
and $j$ exceeds its hop number $H_i$, which can be expressed as

\begin{equation}
{h_{ij}} = \left\{ {\begin{array}{*{20}{c}}
{1,\;{\rm{if}}\;{d_i} > 0\;\&\;j \le {H_i}},\\
{0,\;{\rm{otherwise}}.\hspace{1.3cm}}
\end{array}} \right.\;\forall \;i,j.\hspace{2.2cm}\label{CONS1}
\end{equation}
In a schedule, the traffic demand of each flow $i$ should be accommodated no matter which path its
traffic is transmitted through, which can be expressed as

\begin{equation}\hspace{0.2cm}
\sum\limits_{k = 1}^K \hspace{-0.1cm}{({\delta ^k}  b_{ij}^k  {c_{ij}^b}\hspace{-0.1cm} +
\hspace{-0.1cm}{h_{ij}}{\delta ^k} a_i^k c_i^d)} \left\{ {\begin{array}{*{20}{c}}
{\hspace{-0.2cm} \ge {d_i},{\rm{if}}\;{d_i} > 0\;\&\;j \le {H_i}},\\
{\hspace{-0.2cm}=0,\;{\rm{otherwise}}.\hspace{1.4cm}}
\end{array}\forall \;i,j.} \right.\label{CONS5}
\end{equation}

To avoid frequent beamforming or steering, each link can be activated at most once in a schedule
\cite{mao}. Besides, we assume the traffic of each flow can only be transmitted through its direct
path or its ordinary path. If its traffic is scheduled on the ordinary path, each hop of its
ordinary path should be activated once in one stage. If its traffic is scheduled on the direct
path, its direct link should be activated once in one stage. Thus, the formulated constraint can be
expressed as follows.

\begin{equation}
\sum\limits_{k = 1}^K {(b_{ij}^k + {h_{ij}}a_i^k)} \left\{ {\begin{array}{*{20}{c}}
{ = 1,{\rm{if}}\;{d_i} > 0\;\&\;j \le {H_i}},\\
{ = 0,{\rm{otherwise}}.\hspace{1.3cm}}
\end{array}} \right.\forall\;i,j.\hspace{0.4cm}\label{CONS4}
\end{equation}

Adjacent links cannot be scheduled concurrently in the same stage due to the half-duplex
assumption, which can be formulated as follows.
\begin{equation}
b_{ij}^k + b_{uv}^k \le 1,\;\; {\rm{if}} \;\;I({s_{ij}},{r_{ij}},{{s_{uv}},{r_{uv}}} ) =1;
\;\;\forall \;i,j,u,v,k; \hspace{0.4cm}\label{CONS7}
\end{equation}

\begin{equation}
a_u^k + a_v^k \le 1, \;\;{\rm{if}} \;\;I(s_u^d,r_u^d,s_v^d,r_v^d)=1; \;\;\forall
\;u,v,k;\hspace{1.6cm}\label{CONS8}
\end{equation}

\begin{equation}
a_i^k + b_{uv}^k \le 1,\;\;{\rm{if}} \;\;I(s_i^d,r_i^d,s_{uv},r_{uv})=1. \;\;\forall
\;i,u,v,k.\hspace{1.0cm}\label{CONS9}
\end{equation}

Due to the inherent order of transmissions on the ordinary path of flow $i$, links on the same path
cannot be scheduled concurrently in the same stage, which is formulated as
\begin{equation}
\sum\limits_{j = 1}^{{H_i}} {b_{ij}^k \le 1}. \;\;\;\;\;\forall\;i,k. \hspace{4.8cm}\label{CONS6}
\end{equation}
Besides, the $j$th hop link of the ordinary path of flow $i$ should be scheduled ahead of the
${(j+1)}$th hop link, which can be formulated as
\begin{equation}
\begin{array}{l}\hspace{-0.8cm}
\sum\limits_{k = 1}^{{K^*}} {b_{ij}^k}  \ge \sum\limits_{k = 1}^{{K^*}} {b_{i(j + 1)}^k},\;\;{\rm{if}}\;H_i>1;\\
\hspace{1.0cm}\forall\;i,j = 1 \sim ({H_{i}} - 1), \;{K^*} =1 \sim K.
\end{array}\label{CONS10}
\end{equation}
This constraint is a group of constraints since $K^*$ varies from 1 to $K$.

To enable concurrent transmissions, the SINR of each link in the same stage should be able to
support its transmission rate, which is formulated for links both on the direct path and ordinary
path as follows.
\begin{equation}
\begin{array}{l}\hspace{-0.2cm}
\frac{{{k_0}{P_t}{l_{{s_{ij}},{r_{ij}}}}^{ - \gamma }b_{ij}^k}}{{W{N_0} + \rho \sum\limits_u
{\sum\limits_v {{f_{{s_{uv}},{r_{ij}}}}b_{uv}^k{k_0}{P_t}{l_{{s_{uv}},{r_{ij}}}}^{ - \gamma } +
\rho \sum\limits_p {{f_{s_p^d,{r_{ij}}}}a_p^k{k_0}{P_t}{l_{s_p^d,{r_{ij}}}}^{ - \gamma }} } } }}
\\\hspace{-0.1cm}\ge  MS({c_{ij}^b}) \times b_{ij}^k;\;\;\forall \;i,j,k;
\end{array}\label{CONS11}
\end{equation}

\begin{equation}
\begin{array}{l}\hspace{-0.2cm}
\frac{{{k_0}{P_t}{l_{s_i^d,\;r_i^d}}^{ - \gamma }a_i^k}}{{W{N_0} + \rho \sum\limits_u {\sum\limits_v {{f_{{s_{uv}},\;r_i^d}}b_{uv}^k{k_0}{P_t}{l_{{s_{uv}},\;r_i^d}}^{ - \gamma } + \rho \sum\limits_q {{f_{s_q^d,r_i^d}}a_q^k{k_0}{P_t}{l_{s_q^d,r_i^d}}^{ - \gamma }} } } }}\\
 \hspace{-0.1cm}\ge MS(c_i^d) \times a_i^k.\;\;\forall \;i,k.
\end{array}\label{CONS12}
\end{equation}

Therefore, the problem of optimal scheduling (P1) is formulated as follows.

\begin{equation}
 \min \sum\limits_{k = 1}^K
{{\delta ^k}} ,\label{OBJ} \hspace{5.8cm}
\end{equation}
\hspace{0.45cm}s. t.

\begin{equation}\hspace{-1cm}
b_{ij}^k \in \left\{ {\begin{array}{*{20}{c}}
{\{ 0,1\} ,\;{\rm{if}}\;{d_i} > 0\;\&\;j \le {H_i}},\\
{\{ 0\} ,\;\;\;\;\;\;\;\;\;\;\;\;\;\;\;\;\;\;{\rm{otherwise}};\hspace{0cm}}
\end{array}\forall \;i,j,k} \right.\hspace{0cm}\label{CONS2}
\end{equation}

\begin{equation}\hspace{-2.6cm}
a_i^k \in \left\{ {\begin{array}{*{20}{c}}
{\{ 0,1\} ,\;\;{\rm{if}}\;{d_i} > 0,}\\
{\{ 0\} ,\;\;\;{\rm{otherwise}};\hspace{0.0cm}}
\end{array}\forall\;i,k} \right.\hspace{0cm}\label{CONS3}
\end{equation}
\hspace{0.7cm}Constraints (\ref{CONS1}--\ref{CONS12}).

\vbox{}

In the above formulated problem, constraint (\ref{CONS2}) indicates $b_{ij}^k$ is a binary variable
and is set to 0 if flow $i$ does not have traffic demand, or $j$ exceeds $H_i$. Constraint
(\ref{CONS3}) indicates $a_i^k$ is a binary variable and is set to 0 if flow $i$ does not have
traffic demand.


%

%

%
%
%
%
%
%


\subsection{Problem Reformulation}\label{S4-2}

In Problem P1, we can observe that constraints (\ref{CONS5}), (\ref{CONS11}), and (\ref{CONS12})
have second-order terms and thus are nonlinear constraints. Thus, problem P1 is a mixed integer
nonlinear program (MINLP), which is generally NP-hard. For these second-order terms, we adopt a
relaxation technique, the Reformulation-Linearization Technique (RLT) \cite{RLT, mao_20} to
linearize them. RLT can produce tight linear programming relaxations for an underlying nonlinear
and non-convex polynomial programing problem. RLT applies a variable substitution for each
nonlinear term in the problem to linearize the objective function and the constraints. Nonlinear
implied constraints for each substitution variable are also generated as the products of bounding
terms of the decision variables, up to a suitable order.

For the second order terms, ${\delta ^k}  b_{ij}^k$ and ${\delta ^k} a_i^k$, in constraint
(\ref{CONS5}), we define two substitution variables, $u_{ij}^k={\delta ^k}  b_{ij}^k$ and
$v_i^k={\delta ^k} a_i^k$ as in \cite{mao}. We also define
\begin{equation}
{\delta _{\max }} = \mathop {\max }\limits_{\forall \;i,j = 1 \sim {H_i}} \{ \left\lceil
{{d_i}/c_{ij}^b} \right\rceil ,\left\lceil {{d_i}/c_i^d} \right\rceil |c_{ij}^b \ne 0,c_i^d \ne 0\}
\end{equation}
as the maximum possible number of time slots of a stage. Then we know $0 \le {\delta ^k} \le
{\delta _{\max }} $. Since $0 \le b_{ij}^k \le 1$, we can obtain the \emph{RLT bound-factor product
constraints} for $u_{ij}^k$ as follows \cite{mao}.

\begin{equation}
\left\{ {\begin{aligned}&{
{u_{ij}^k \ge 0}},\\
&{{\delta _{\max }}   b_{ij}^k - u_{ij}^k \ge 0},\\
&{{\delta ^k} - u_{ij}^k \ge 0},\\
&{ {\delta _{\max }} - {\delta ^k} - {\delta _{\max }}  b_{ij}^k + u_{ij}^k \ge 0 };
\end{aligned}\;\;\;{\rm{for}}\;\;\;{\rm{all}}\;\;i,j,k}. \right.\\
\label{RLT 1}
\end{equation}

For $v_i^k$, we can obtain its \emph{RLT bound-factor product constraints} in a similar way. The
RLT procedures for the second-order terms in constraints (\ref{CONS11}) and (\ref{CONS12}) are
similar and thus omitted. After substituting these substitution variables into their responding
constraints, we can reformulate problem P1 into a mixed integer linear program (MILP) as
follows.

\begin{equation}
 \min \sum\limits_{k = 1}^K
{{\delta ^k}} ,\label{OBJ2} \hspace{7.3cm}
\end{equation}
\hspace{0cm}s. t.

\begin{equation}\hspace{0.0cm}
\sum\limits_{k = 1}^K \hspace{-0.1cm}{(u_{ij}^k  {c_{ij}^b}\hspace{-0.1cm} +
\hspace{-0.1cm}{h_{ij}}v_i^k c_i^d)} \left\{ {\begin{array}{*{20}{c}}
{\hspace{-0.2cm} \ge {d_i},{\rm{if}}\;{d_i} > 0\;\&\;j \le {H_i}},\\
{\hspace{-0.2cm}=0,\;{\rm{otherwise}};\hspace{1.4cm}}
\end{array}\forall \;i,j} \right.\label{CONS5_RLT}
\end{equation}

\hspace{0.15cm}Constraint (\ref{RLT 1}) and generated RLT bound-factor product \\

\hspace{0.15cm}constraints for $v_i^k$.\\

\hspace{0.15cm}Constraints (\ref{CONS11}) and (\ref{CONS12}) after the RLT procedure and \\

\hspace{0.15cm}generated RLT bound-factor product constraints.\\

\hspace{0.15cm}Constraints (\ref{CONS1}), (\ref{CONS4}--\ref{CONS10}) and
(\ref{CONS2}--\ref{CONS3}).

\vbox{}


We consider the example in Fig. \ref{scenario}. The traffic demand vector $\textbf{d}$, the
transmission rate matrix for the ordinary paths as $\textbf{C}^b$, and the transmission rate vector
for the direct paths $\textbf{c}^d$ are the same as those in Section \ref{S3-2}. We also assume for
any two nonadjacent links $(s_i,r_i)$ and $(s_j,r_j)$, $f_{{s_i},{r_j}}$ is equal to 0. Then we
solve problem (\ref{OBJ2}) using an open-source MILP solver, YALMIP \cite{yalmip}. The path
selection scheme is that flow A $\to$ B goes through its ordinary path, and the other three flows
go through their direct paths. The optimal schedule has three stages, and is already illustrated in
Fig. \ref{frame}.

The formulated MILP is NP-hard. The number of constraints is $\mathcal{O}(({FH_{max}})^2K)$, where
$H_{max}$ is the maximum number of hops of ordinary paths of flows. The number of decision
variables is $\mathcal{O}((FH_{max})^2K)$. Using the branch-and-bound algorithm will take
significantly long computation time \cite{mao}, and it is unacceptable for practical mmWave cells
where the duration of one time slot is only a few microseconds. Therefore, we need heuristic
algorithms with low computational complexity to obtain near-optimal solutions in practice.

\section{The D2DMAC Scheme}\label{S5}

To solve the formulated MILP, we should consider two steps. At the first step, we should select an
appropriate transmission path, the direct path or ordinary path, for each flow. Specially, direct transmissions should be enabled if there are direct paths of high channel quality. At the second
step, we should exploit concurrent transmissions fully to improve transmission
efficiency as much as possible. Following the above ideas, in this section, we propose a path
selection criterion to decide transmission paths for flows, and a transmission scheduling algorithm
to compute schedules to accommodate the traffic demand of flows with as few time slots as possible.


\subsection{The Path Selection Criterion}\label{S5-1}

Intuitively, for each flow, if its direct path has high channel quality, we should enable its
direct transmission other than transmissions through its ordinary path. For a transmission path $p$
of $h$ hops, we define its transmission capability as

\begin{equation}
A(p) = \frac{1}{{\sum\limits_{j = 1}^h {\frac{1}{{{c_j}}}} }},
\end{equation}
where $c_j$ is the transmission rate of the $j$th hop on $p$. Therefore, for each flow, we should
choose the path with higher transmission capability between its direct path and ordinary path. The
path selection criterion can be expressed as

\begin{equation}
\left\{ {\begin{array}{*{20}{c}}
{{\rm{if}}\;\frac{{A(p_i^d)}}{{A(p_i^b)}} \ge \beta ,\;{\rm{choose}}\;p_i^d;}\\
{{\rm{otherwise}},\;\;\;\;\;{\rm{choose}}\;p_i^b.}
\end{array}} \right.\;\;\;\;\forall \;i.
\end{equation}
For flow $i$, ${p_i^d}$ denotes its direct path, and ${p_i^b}$ denotes its ordinary path. $\beta$
is the path selection parameter, which is larger than or equal to 1. The smaller $\beta$, the
higher the priority of direct transmissions between devices.

For the example in section \ref{S3-2}, we apply the path selection criterion with $\beta$ equal to
2, and the result is that flow A $\to$ B will be transmitted through its ordinary path, and the
other three flows will be transmitted through their direct paths, which is the same as the optimal
solution in Section \ref{S4-2}.


\subsection{The Transmission Scheduling Algorithm}\label{S5-2}

After selecting transmission paths for flows, we propose a heuristic transmission scheduling algorithm to
accommodate the traffic demand of flows with as few time slots as possible by exploiting spatial reuse fully. Since adjacent links cannot be scheduled concurrently in the same stage, links scheduled in the
same stage should be a matching, and the maximum number of links in
the same stage can be inferred to be $\left\lfloor {n/2} \right\rfloor $ \cite{mao}. On the other hand, due to the
inherent order of transmissions on each path, preceding hops should be scheduled ahead of
subsequent hops since nodes behind on the path are able to relay the packets after receiving the packets from preceding nodes. Borrowing the design ideas of greedy coloring (GC) algorithm in FDMAC \cite{mao},
we perform edge coloring on the first unscheduled hops of paths for all flows. After each coloring,
we remove scheduled hops, and the set of first unscheduled hops is updated since the hops after
these scheduled hops become the first unscheduled hops now. Our algorithm schedules the
first unscheduled hops of flows iteratively into each stage in non-increasing order of weight with the conditions for
concurrent transmissions satisfied. To maximize spatial reuse, our
algorithm terminates scheduling in each stage when no possible link can be scheduled into this stage any more, i.e., when all
possible links are examined, or the number of links in this stage reaches $\left\lfloor {n/2}
\right\rfloor $. Our algorithm carries out this process iteratively until all hops of flows are
properly scheduled.

For each
flow $i$, we denote its selected transmission path by the path selection criterion by $p_i$. We denote the number of nodes including APs and WNs by $n$, and the set of
selected paths for all flows by $P$, which includes the transmission path $p_i$ of each flow $i$.
The set of hops in $P$ is denoted by $H$. For each
hop of each flow's transmission path, we define the number of time slots to accommodate its traffic
as the weight of this hop. For the $j$th hop of flow $i$, its weight is denoted by $w_{ij}$, and we
denote this link by $(s_{ij},r_{ij})$ with transmission rate $c_{ij}$. The sequence number of the
first unscheduled hop on path $p_i$ is denoted by $F_u(p_i)$. In the $t$th stage, the set of paths
that are not visited yet is denoted by $P_u^{t}$. We denote the set of directional links of the
$t$th stage by $H^t$, and the set of vertices of the links in $H^t$ by $V^t$.


The pseudo-code of our transmission scheduling algorithm is presented in Algorithm \ref{alg:BAD}.
In the initialization, we obtain the set of transmission paths of flows selected by the path selection criterion,
$P$. From $P$, we also obtain the set of hops, $H$, and their weights. Since we should start
scheduling the first hop of each path, $F_u(p_i)$ for each path $p_i$ is set to 1. As indicated in
line 1, our algorithm iteratively schedules each hop of $H$ into each stage until all hops in $H$
are scheduled successfully. In the beginning of each stage, all paths in $P$ are unvisited, as in
line 4. In each stage, our algorithm iteratively visits each possible hop until all possible hops
are visited, or the number of links in the stage reaches $\left\lfloor {n/2} \right\rfloor $, as
indicated by lines 5--24. In line 6, our algorithm obtains the set of first unscheduled hops of unvisited
paths, denoted by $H_u^t$, and obtains the hop in $H_u^t$ with the largest weight in
line 7, $(s_{iF_u(p_i)}, r_{iF_u(p_i)})$. Then our algorithm checks whether this hop is adjacent to the links already in this stage,
as indicated by line 8. If it is not, this hop will be selected as the candidate link, and be added
to this stage in lines 9--10. Then the condition for concurrent transmissions of each link in this
stage is checked, as indicated by lines 11--16. If any of links in this stage cannot support its
transmission rate, the candidate link will be removed from this stage, as indicated by lines 13--14
and 20--21. If the concurrent transmission conditions of links in this stage are satisfied, the
newly added link will be removed from $H$, as indicated by line 17. The number of time slots of
this stage is updated to accommodate the traffic of this newly added link, as indicated by line 18.
The sequence number of first unscheduled hop of the path, where the newly added link is, will
increase by 1, as indicated by line 19. The path where this candidate link is will be removed from
the set of unvisited paths whether this link is added to this stage or not, as indicated by line
23. The scheduling results for each stage are outputted in line 25.

We apply this scheduling algorithm to the example in Section \ref{S4-2}, and obtain the schedule as
follows: in the first stage, the first hop A $\to$ AP2 of flow A $\to$ B, link B $\to$ C, and link
D $\to$ AP1 transmit for three time slots; in the second stage, the second hop AP2 $\to$ AP3 of
flow A $\to$ B transmits for two time slots, and link AP1 $\to$ B for three time slots; in the
third stage, the third hop AP3 $\to$ B of flow A $\to$ B transmits for three time slots. The total
number of time slots of this schedule is the same as that obtained by using optimization softwares.
However, our algorithm has the computational complexity of $\mathcal{O}(n^4)$, which is a
pseudo-polynomial time solution and can be implemented in practice.



\begin{algorithm}[htbp]
\caption{The Transmission Scheduling Algorithm} \label{alg:BAD}
\begin{algorithmic}[1]
\REQUIRE ~~\\
 Input: the set of selected paths of all flows, $P$; \\
 Obtain the set of hops in $P$, denoted by $H$; \\
 Obtain the weight of each hop $(s_{ij},r_{ij})\in H$, $w_{ij}$;\\
 Set $F_u(p_i)=1$ for each $p_i\in P$; $t$=0;
\ENSURE ~~\\
\WHILE {($|H| > 0$)}
\STATE  $t$=$t$+1;  \\
\STATE  Set ${V^t} = \emptyset $, ${H^t} = \emptyset $, and $\delta^{t}=0$; \\
\STATE Set $P_u^t$ with $P_u^t = P$; \\
\WHILE {($|{P_u^t}|>0$ and $|{H^t}| < \left\lfloor {n/2} \right\rfloor $)}
\STATE Obtain the set of first unscheduled hops of $P_u^t$, $H_u^t$;
\STATE Obtain the hop $(s_{iF_u(p_i)}, r_{iF_u(p_i)})\in H_u^t$ with the largest weight;\\
 \IF {($s_{iF_u(p_i)} \notin {V^t}$\;and\;$ r_{iF_u(p_i)} \notin {V^t}$)}
\STATE  ${H^t} = {H^t} \cup \{ (s_{iF_u(p_i)}, r_{iF_u(p_i)})\}$; \\\STATE ${V^t} = {V^t} \cup \{ s_{iF_u(p_i)},r_{iF_u(p_i)}\} $;\\
 \FOR {each link $(s_{ij},r_{ij})$ in ${H^t}$}
\STATE  Calculate the SINR of link $(s_{ij},r_{ij})$, $SIN{R_{s_{ij}r_{ij}}}$\\
\IF {($SIN{R_{s_{ij}r_{ij}}}<MS({c_{ij}})$)}
\STATE  Go to line 20\\
\ENDIF \ENDFOR
\STATE    $H=H-\{(s_{iF_u(p_i)}, r_{iF_u(p_i)})\}$; \\
\STATE $\delta^t={\rm{max}}\{\delta^t, w_{iF_u(p_i)}\}$;\\
\STATE $F_u(p_i)=F_u(p_i)+1$; Go to line 22\\
\STATE   ${H^t} = {H^t} - \{ (s_{iF_u(p_i)}, r_{iF_u(p_i)})\}$; \\ \STATE ${V^t} = {V^t} - \{ s_{iF_u(p_i)},r_{iF_u(p_i)}\}$;\\
\ENDIF\\
\STATE $P_u^t = P_u^t - p_i$;\\
\ENDWHILE\\

 \STATE Output $H^t$ and ${\delta ^t}$;\\
\ENDWHILE

\end{algorithmic}
\end{algorithm}

\begin{remark}

In D2DMAC, only a small amount of control information should be transmitted from the central controller (APs and WNs) to the APs and WNs (the central controller). In the scheduling phase, a small amount of control data regarding traffic demand of WNs and APs should be collected by APs and transmitted to the central controller through the backhaul network. As in \cite{mao}, with the Gbps transmission rate, the traffic demand of WNs can be collected by APs in a few time slots. Similarly, with the optimized backhaul network, the central controller is able to collect the traffic demand information from APs in a few time slots. After schedule computation, the central controller should push the schedule to APs and WNs. In the same manner, the central controller is able to push a small amount of control data regarding the schedule to APs and WNs in a few time slots. The number of time slots in the traffic polling or schedule pushing depends on the number of APs and WNs and the transmission rates of links between the central controller and APs and links between APs and WNs.

\end{remark}

\begin{remark}
Retransmission is an important mechanism to improve the robustness of transmission. Retransmissions can be enabled in D2DMAC. In the traffic polling part in the scheduling phase, the receivers (APs or WNs) are able to collect the failed transmissions in the last frame, and if one failed packet needs to be retransmitted and its retransmission times do not exceed the maximum limit, the receiver will report the failed packet to the central controller. Then the central controller will schedule the retransmission of the failed packet in the current frame.

\end{remark}

\section{Performance Analysis}\label{S5-i}

The conditions for concurrent transmissions determine the links scheduled concurrently in the same stage in D2DMAC. Thus, they have an important impact on the scheduling efficiency of D2DMAC, and we should study them further for a
better understanding of D2DMAC. In this section, we give a theoretical performance analysis by
presenting the conditions for the concurrent transmissions in the same stage in D2DMAC. In the next section, we demonstrate the superior performance of D2DMAC by extensive simulations under various traffic patterns, and comparison with the Optimal Solution and other existing protocols.

Suppose there are $M$ nonadjacent links, $({s_1},{r_1})$, $({s_2},{r_2})$, ... , $({s_M},{r_M})$,
where $M$ is less than or equal to $\left\lfloor {n/2} \right\rfloor$ \cite{mao}. To enable concurrent
transmissions of the $M$ links, the SINR of each link should be larger than or equal to the minimum
SINR to support its transmission rate. For link $({s_1},{r_1})$, its SINR should meet
\begin{equation}
\frac{{{k_0}{P_t}{l_{{s_1}{r_1}}}^{ - \gamma }}}{{W{N_0} + \rho \sum\limits_{m = 2}^M
{{f_{{s_m},{r_1}}}{k_0}{P_t}{l_{{s_m}{r_1}}}^{ - \gamma }} }} \ge MS({c_{{s_1}{r_1}}}).
\end{equation}
It can be converted to
\begin{equation}
 \sum\limits_{m = 2}^M
{{f_{{s_m},{r_1}}}{l_{{s_m}{r_1}}}^{ - \gamma }}  \le (\frac{{{k_0}{P_t}{l_{{s_1}{r_1}}}^{ - \gamma
}}}{{MS({c_{{s_1}{r_1}}})}} - W{N_0})/(\rho {k_0}{P_t}),\label{concurrent_condition}
\end{equation}
which can be regarded as the \emph{Spatial Reuse Region} of link $({s_1},{r_1})$ and is denoted as
$SR{R_1}$. If ${{l_{{s_2}{r_1}}}}$, ${{l_{{s_3}{r_1}}}}$, ... , ${{l_{{s_M}{r_1}}}}$ are
independent and identically distributed random variables with probability density function $f(x)$,
then the probability of link $({s_1},{r_1})$ to support transmission rate ${c_{{s_1}{r_1}}}$ with
the interference from $({s_2},{r_2})$, ... , $({s_M},{r_M})$ considered is
\begin{equation}
{B_{{s_1}{r_1}}} = \int  \cdots  \int_{SR{R_1}} {f({l_{{s_2}{r_1}}}) \cdots
f({l_{{s_M}{r_1}}})d{l_{{s_2}{r_1}}} \cdots d{l_{{s_M}{r_1}}}}.
\end{equation}
Similarly, the \emph{Spatial Reuse Regions} of links $({s_2},{r_2})$, ... , $({s_M},{r_M})$ can
also be obtained and are denoted as $SR{R_2}$, ... , $SR{R_M}$. The probability of these links to
support their transmission rates can also be obtained and is denoted as ${B_{{s_2}{r_2}}}$,
${B_{{s_3}{r_3}}}$, ... , ${B_{{s_M}{r_M}}}$. Therefore, the probability of concurrent
transmissions of $({s_1},{r_1})$, $({s_2},{r_2})$, ... , $({s_M},{r_M})$ can be calculated as
\begin{equation}
{B_{{s_1}{r_1},{s_2}{r_2}, \cdots ,{s_M}{r_M}}} = {B_{{s_1}{r_1}}}{B_{{s_2}{r_2}}} \cdots
{B_{{s_M}{r_M}}}.
\end{equation}

For link $({s_1},{r_1})$, if there are ${F_{{s_1}{r_1}}}$ links that have interference to link
$({s_1},{r_1})$, i.e. $\sum\limits_{m = 2}^M {{f_{{s_m},{r_1}}}}  = {F_{{s_1}{r_1}}}$, then a
sufficient condition for $({s_1},{r_1})$ to support transmission rate ${c_{{s_1}{r_1}}}$ can be
inferred from (\ref{concurrent_condition}) and is illustrated as follows. For each link
$({s_m},{r_m})$ of the ${F_{{s_1}{r_1}}}$ links, if the distance between its transmitter $s_m$ and
node $r_1$, ${l_{{s_m}{r_1}}}$ meets

\begin{equation}
{l_{{s_m}{r_1}}} \ge {(\rho {k_0}{P_t}{F_{{s_1}{r_1}}})^{1/\gamma
}}/{(\frac{{{k_0}{P_t}{l_{{s_1}{r_1}}}^{ - \gamma }}}{{MS({c_{{s_1}{r_1}}})}} - W{N_0})^{1/\gamma
}},\label{radius}
\end{equation}
the SINR of $({s_1},{r_1})$ will be able to support its transmission rate, and the right side of
(\ref{radius}) can be seen as the \emph{interference radius} of link $({s_1},{r_1})$. The distances
between the transmitters of the ${F_{{s_1}{r_1}}}$ links and receiver $r_1$ should be larger than
or equal to the interference radius to support the transmission rate of $({s_1},{r_1})$. Similarly,
the interference radiuses of other links can also be calculated.

\begin{figure}[htbp]
\begin{minipage}[t]{0.5\linewidth}
\centering
\includegraphics[width=1\columnwidth,height=1.3in]{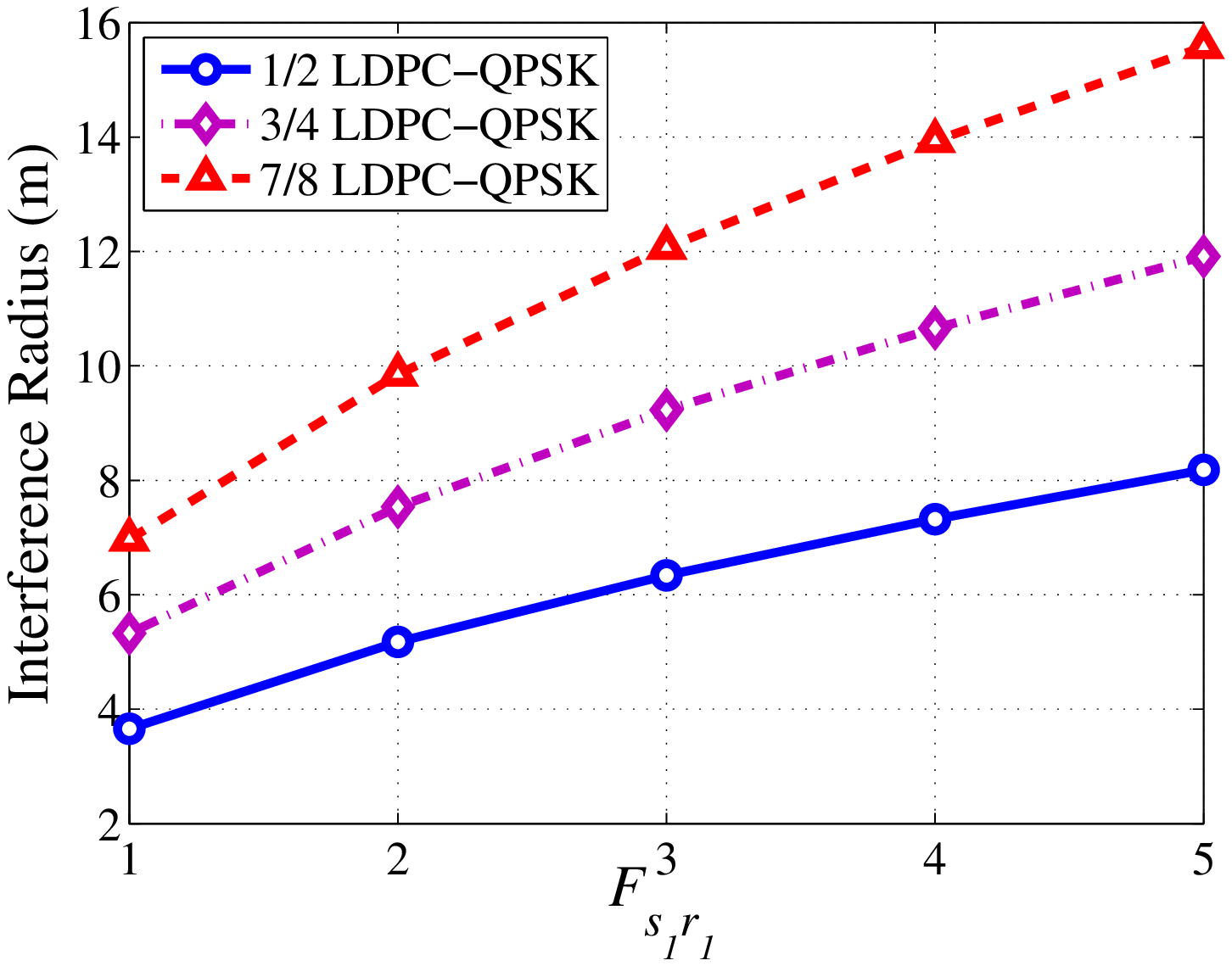}
\centerline{\small (a) Different MCSs}
\end{minipage}%
\begin{minipage}[t]{0.5\linewidth}
\centering
\includegraphics[width=1\columnwidth,height=1.3in]{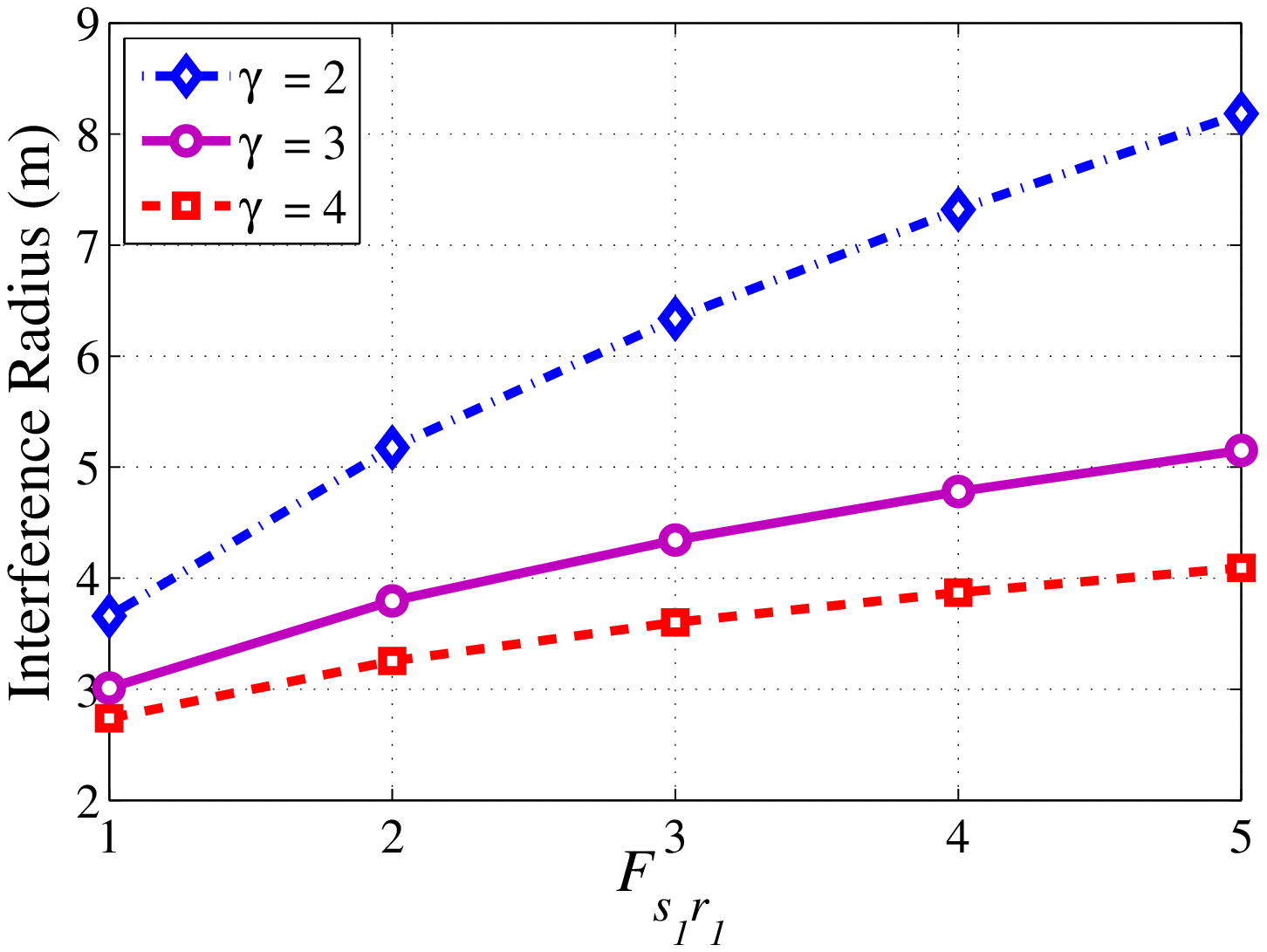}
\centerline{\small (b) Different path loss exponents}
\end{minipage}%
\caption{Interference radii under different numbers of interference links.}
\label{fig:infer_radii} 
\vspace*{-3mm}
\end{figure}

In Fig. \ref{fig:infer_radii}, we plot the interference radii of link $({s_1},{r_1})$ under
different numbers of interference links. The bandwidth $W$ is set to 1760 MHz, and $l_{{s_1}{r_1}}$
is set to 2 m. $\rho$ is set to 1, and the transmit power $P_t$ is set to 0.1 mW. Other parameters
are the same as those in Table I of Ref. \cite{Qiao}. In Fig. \ref{fig:infer_radii} (a), we compare
three modulation and coding schemes in IEEE 802.11ad, QPSK with LDPC codes of coding rate 1/2, 3/4,
and 7/8 respectively. According to their BER performance in Ref. \cite{mcs}, their minimum SINRs to
support transmission rates of 1760 Mbps, 2640 Mbps, and 3080 Mbps are 5 dB, 8 dB, and 10 dB
respectively. The path loss exponent is set to 2. From the results, we can observe that the
interference radius increases with the number of interference links, ${F_{{s_1}{r_1}}}$. The more
interference sources, the farther they should be located relative to receiver $r_1$ to reduce
interference. On the other hand, the interference radius also increases with the minimum SINR to
support ${c_{{s_1}{r_1}}}$, $MS({c_{{s_1}{r_1}}})$. When ${c_{{s_1}{r_1}}}$ increases, higher SINR
is required, and the interference sources should be located farther relative to receiver $r_1$ to
avoid severe interference.

In Fig. \ref{fig:infer_radii} (b), we compare the interference radii of QPSK with LDPC codes of
coding rate 1/2 under different path loss exponents. As we can see, the interference radius
decreases with the path loss exponent. With a higher path loss exponent, the interference power
decreases more rapidly, especially with more interference sources, and thus interference sources
can be located nearer to receiver $r_1$.

\begin{figure}[htbp]
\begin{minipage}[t]{0.5\linewidth}
\centering
\includegraphics[width=1\columnwidth,height=1.3in]{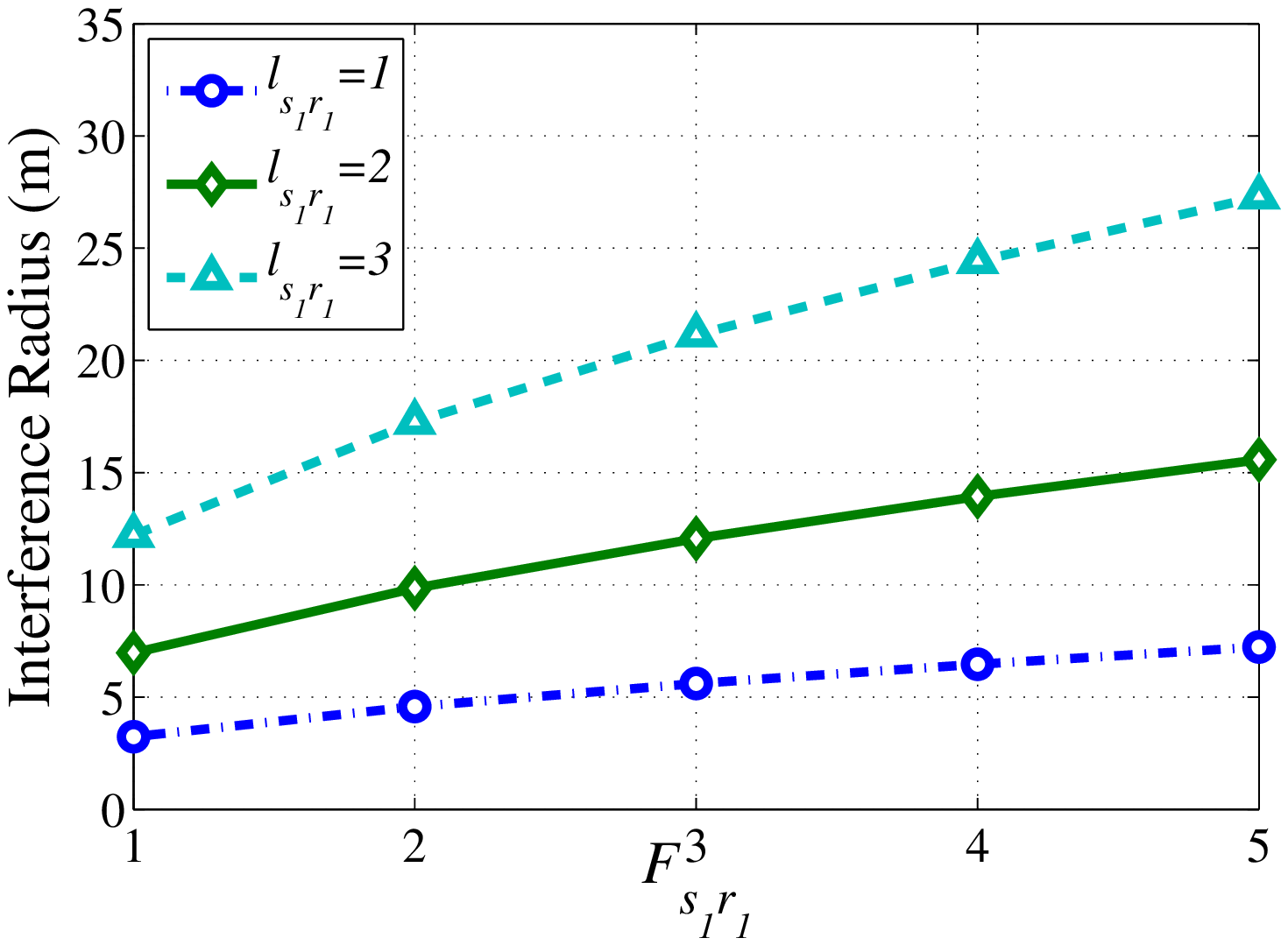}
\centerline{\small (a) Different link lengths }
\end{minipage}%
\begin{minipage}[t]{0.5\linewidth}
\centering
\includegraphics[width=1\columnwidth,height=1.3in]{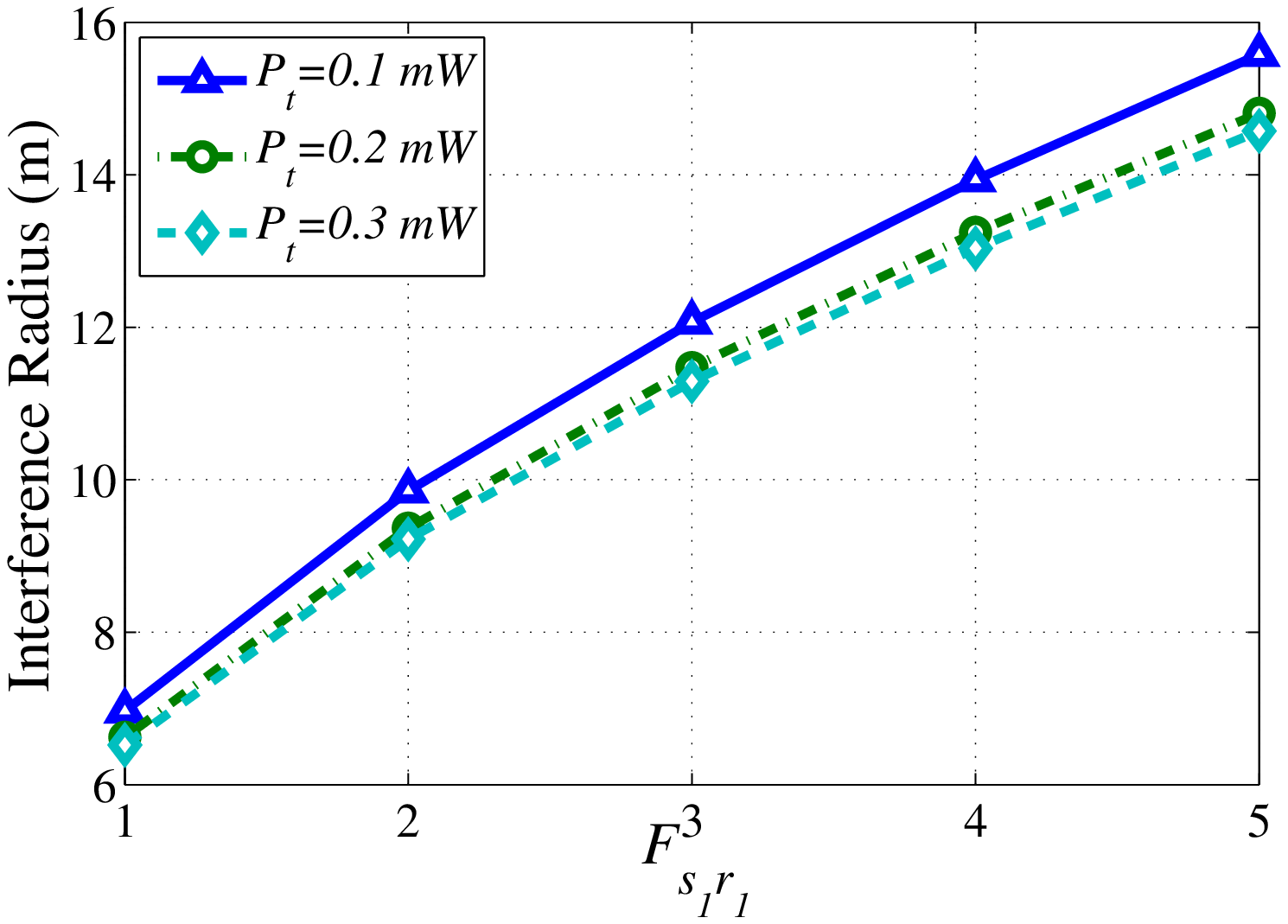}
\centerline{\small (b) Different transmit power}
\end{minipage}%
\caption{Interference radii under different numbers of interference links.}
\label{fig:infer_radii_2} 
\vspace*{-3mm}
\end{figure}

In Fig. \ref{fig:infer_radii_2}, we compare the interference radii of QPSK with LDPC codes of
coding rate 7/8 under different link lengths and transmit power. From Fig. \ref{fig:infer_radii_2}
(a), we can observe the interference radius increases with the link length $l_{{s_1}{r_1}}$
significantly. The longer the link length, the less the signal power received. Thus, the interference
sources should be located farther relative to receiver $r_1$. Therefore, in the user densely distributed scenario, the link length is shorter on average, and the interference radius will also be shorter, which indicates the concurrent transmission conditions can be satisfied more easily in this case. From Fig. \ref{fig:infer_radii_2}
(b), we can observe the interference radius decreases with the transmit power. With more transmit
power, more signal power is received, and thus interference sources can be located nearer to
receiver $r_1$.

\section{Performance Evaluation}\label{S6}

In this section, we give an extensive performance evaluation for our proposed joint scheduling
scheme, D2DMAC, under various traffic patterns. Specifically, we compare its performance with the
optimal solution and other protocols, and analyze its performance with different path selection
parameters.





\subsection{Simulation Setup}\label{S6-1}


In the simulation, we consider a typical dense deployment of small cells in the 60 GHz band
\cite{dense cells}, where nine APs are uniformly distributed in a square area of 50 m $ \times$
50 m, and the gateway is located at the center point. There are 30 WNs uniformly distributed in this
area, and each WN is associated with the nearest AP. We adopt the simulation parameters in Table II
of Ref. \cite{MRDMAC}, which are listed in Table \ref{tab:simulation_parameter}. We set the
duration of one time slot to 5 $\mu s$ and the size of data packets to 1000 bytes \cite{mao}. We set three
transmission rates, 2 Gbps, 4 Gbps, and 6 Gbps according to the distances between devices. For
links of the backhaul network, we set their transmission rates to 6 Gbps due to their better
channel conditions. With a transmission rate of 2 Gbps, a packet can be transmitted in a time slot
\cite{mao}. The AP can access $\left\lfloor
{\frac{{{T_{slot}}}}{{{T_{ShFr}} + 2 \cdot {T_{SIFS}} + {T_{ACK}}}}}
\right\rfloor $ nodes in one time slot in \cite{mao}. With a small cell of 10 nodes,
the AP is able to complete the traffic demand polling stage or the schedule
pushing in one time slot. Similarly, with the optimized backhaul network, the control information transmission between the central controller and APs can also be completed in a few time slots. Thus, for the simulated network, the central controller is able to complete the polling of
traffic or schedule pushing in a few time slots. Generally, it takes only a few time slots for the
central controller to complete schedule computation. The simulation length is set to 0.5 seconds.
The delay threshold is set to $10^4$ time slots, and packets with delay larger than the threshold
will be discarded. Initially, each flow has a few packets randomly generated to transmit. In the simulation, we assume nonadjacent links are able to be scheduled for concurrent transmissions. Since retransmission is not the focus of this paper, we do not consider it in the simulations.

\begin{table}
\begin{center}
\caption{Simulation Parameters}
\def \temptablewidth {0.9\textwidth}
\begin{tabular}{ccc}
\hline
\textbf{Parameter}&\textbf{Symbol}&\textbf{Value}\\
\hline
PHY data rate & R & 2Gbps, 4Gbps, 6 Gbps \\
Propagation delay&${\delta _p}$& 50ns\\
Slot Duration & $T_{slot}$ & 5 $\mu s$\\
PHY overhead& ${T_{PHY}}$ & 250ns\\
Short MAC frame Tx time& ${T_{ShFr}}$& ${T_{PHY}}$+14*8/R+${\delta _p}$\\
Packet transmission time&${T_{packet}}$& 1000*8/R\\
SIFS interval&${T_{SIFS}}$& 100ns\\
ACK Tx time&${T_{ACK}}$&${T_{ShFr}}$\\
\hline
\end{tabular}
\label{tab:simulation_parameter}
\end{center}
\end{table}

In the simulation, we set two kinds of traffic modes, the Poisson Process and Interrupted Poisson
Process (IPP).

\subsubsection {\textbf{Poisson}} packets arrive at each flow following the Poisson Process with arrival
rate $\lambda $. The traffic load in Poisson traffic, denoted by ${T_l}$, is defined as
 \begin{equation}
{T_l} = \frac{{\lambda  \times L \times N}}{R}, \label{Tl_1}
\end{equation}
where $L$ is the size of data packets, $N$ is the number of flows, and $R$ is set to 2 Gbps.

\subsubsection {\textbf{IPP}} packets arrive at each flow following an Interrupted Poisson
Process (IPP). The parameters are ${{\lambda _1}}$, ${{\lambda _2}}$, ${{p_1}}$, and ${{p_2}}$. The
arrival intervals obey the second-order hyper-exponential distribution with a mean of
\begin{equation}
E(X) = \frac{{{p_1}}}{{{\lambda _1}}} + \frac{{{p_2}}}{{{\lambda _2}}}.
\end{equation}
The traffic load ${T_l}$ in this mode is defined as
 \begin{equation}
{T_l} = \frac{{L \times N}}{{E(X) \times R}}.\label{Tl_2}
\end{equation}

We evaluate the system performance by the following four metrics. To explain them clearly, we
denote the set of packets transmitted in the simulation for each flow $i$ as $S_i$. For each packet
$e$ in $S_i$, we denote its delay in units of time slots as $y_e$. We also denote the delay
threshold as $TH$.

1) \textbf{Average Transmission Delay:} The average transmission delay of received packets from all
flows; we evaluate it in units of time slots, which can be expressed as

\begin{equation}
{\rm{Average \;Transmission \;Delay}} = \frac{{\sum\limits_{i = 1}^N {\sum\limits_{e \in {S_i}}^{}
{{y_e}} } }}{{\sum\limits_{i = 1}^N {\left| {{S_i}} \right|} }}.
\end{equation}

2) \textbf{Network Throughput:} The total number of successful transmissions of all flows until the
end of simulation. For each packet, if its delay is less than or equal to the threshold, it will be
counted as a successful transmission. With fixed simulation length and packet size, it shows
throughput performance well and can be expressed as

\begin{equation}
{\rm{Network\; Throughput}} = \sum\limits_{i = 1}^N {\left| {\{ e|e \in {S_i},\;{y_e} \le TH\}}
\right|}.
\end{equation}

3) \textbf{Average Flow Delay:} The average transmission delay of some flows, which may be transmitted though ordinary paths or direct paths. In the simulation, we
investigate two cases. In the first case, we evaluate the average transmission delay of flows
between WNs. We denote the set of flows between WNs as ${\rm{B_W}}$. Then Average Flow Delay in
this case can be expressed as
\begin{equation}
{\rm{Average \;Flow \;Delay_1}} = \frac{{\sum\limits_{i \in {\rm{B_W}}}^{} {\sum\limits_{e \in
{S_i}}^{} {{y_e}} } }}{{\sum\limits_{i \in {\rm{B_W}}}^{} {\left| {{S_i}} \right|} }}.
\end{equation}

In the second case, we evaluate the average transmission delay of flows from or to the Internet.
With the set of flows from or to the Internet denoted as ${\rm{I_N}}$, Average Flow Delay in the
second case can be expressed as

\begin{equation}
{\rm{Average \;Flow \;Delay_2}} = \frac{{\sum\limits_{i \in {\rm{I_N}}}^{} {\sum\limits_{e \in
{S_i}}^{} {{y_e}} } }}{{\sum\limits_{i \in {\rm{I_N}}}^{} {\left| {{S_i}} \right|} }}.
\end{equation}

4) \textbf{Flow Throughput:} The number of successful transmissions achieved by each flow until the
end of simulation. In the simulation, we also investigate two cases. In the first case, we evaluate
the average flow throughput of flows between WNs, which can be expressed as
\begin{equation}
{\rm{Flow \;Throughpu}}{{\rm{t}}_1}{\rm{ = }}\frac{{\sum\limits_{i \in {{\rm{B_W}}}}^{} {\left| {\{
e|e \in {S_i},\;{y_e} \le TH\}} \right|} }}{{\left| {{{\rm{B_W}}}} \right|}}.
\end{equation}

In the second case, we evaluate the average flow throughput of flows from or to the Internet, which
can be expressed as
\begin{equation}
{\rm{Flow \;Throughpu}}{{\rm{t}}_2}{\rm{ = }}\frac{{\sum\limits_{i \in {{\rm{I_N}}}}^{} {\left| {\{
e|e \in {S_i},\;{y_e} \le TH\}} \right|} }}{{\left| {{{\rm{I_N}}}} \right|}}.
\end{equation}

In the simulation, we compare the performance of D2DMAC with the following three benchmark schemes:

1) \emph{\textbf{ODMAC}}: In ODMAC, device-to-device transmissions are not enabled, and all flows
are transmitted through their ordinary paths. The scheduling algorithm of ODMAC is the same as
D2DMAC. This benchmark scheme represents the current state-of-the-art work in terms of scheduling
the access or backhaul without considering the device-to-deivce
transmissions\cite{mao,60GHz-backhaul-1, 60GHz-backhaul-4}.

2) \textbf{\emph{RPDMAC}}: RPDMAC selects the transmission path for each flow randomly from its
direct path and ordinary path. Its scheduling algorithm is the same as D2DMAC. Thus, RPDMAC is a
good benchmark scheme to show the advantages of the path selection criterion in D2DMAC.


3) \textbf{\emph{FDMAC-E}}: FDMAC-E is an extension of FDMAC \cite{mao}, and to the best of our knowledge, FDMAC achieves highest efficiency in terms of spatial reuse. In FDMAC-E, the transmission path is selected the same as D2DMAC with the path selection
parameter $\beta$ equalling to 2. However, in order to show the role of backhaul optimization, the access links and backhaul links are separately scheduled in FDMAC-E. The access links from WNs to APs are scheduled by the greedy coloring (GC) algorithm of FDMAC in \cite{mao}. The backhaul links on the transmission path are scheduled by the serial TDMA. The access links from APs to WNs are also scheduled by the GC algorithm of FDMAC.



\subsection{Comparison with the Optimal Solution}

We first compare D2DMAC with the optimal solution of the MILP problem, where the path selection
parameter $\beta$ is set to 2. Since obtaining the optimal solutions takes long time, we simulate a scenario of nine cells with ten users and ten flows.
The simulation length is set to 0.025 seconds, and the delay threshold is set to 50 time slots.
The traffic load is defined the same as in Section \ref{S6-1} with $N$ equal to 10.





\begin{figure}[htbp]
\begin{minipage}[t]{0.5\linewidth}
\centering
\includegraphics[width=1\columnwidth,height=1.3in]{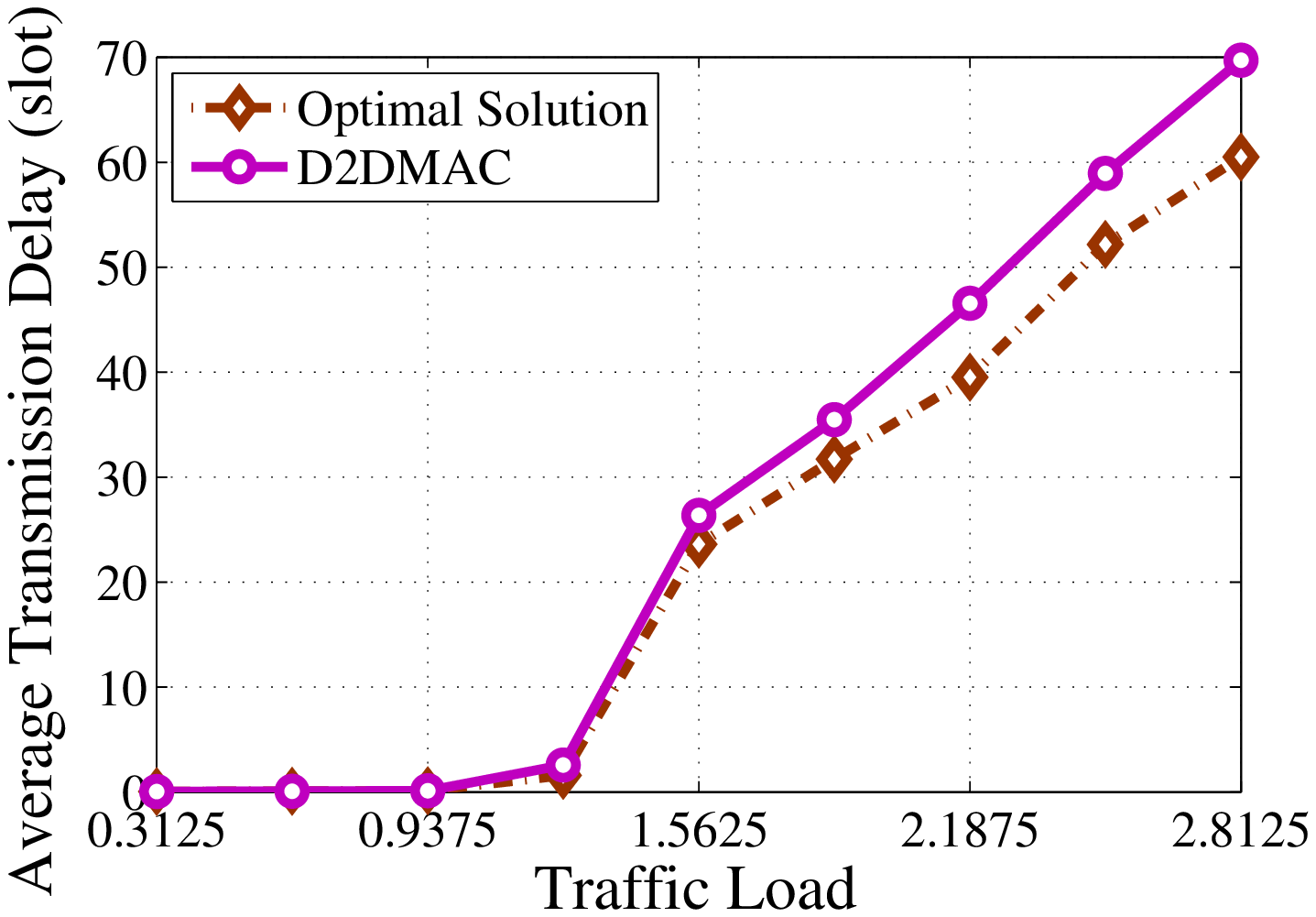}
\centerline{\small (a) Average transmission delay}
\end{minipage}%
\begin{minipage}[t]{0.5\linewidth}
\centering
\includegraphics[width=1\columnwidth,height=1.3in]{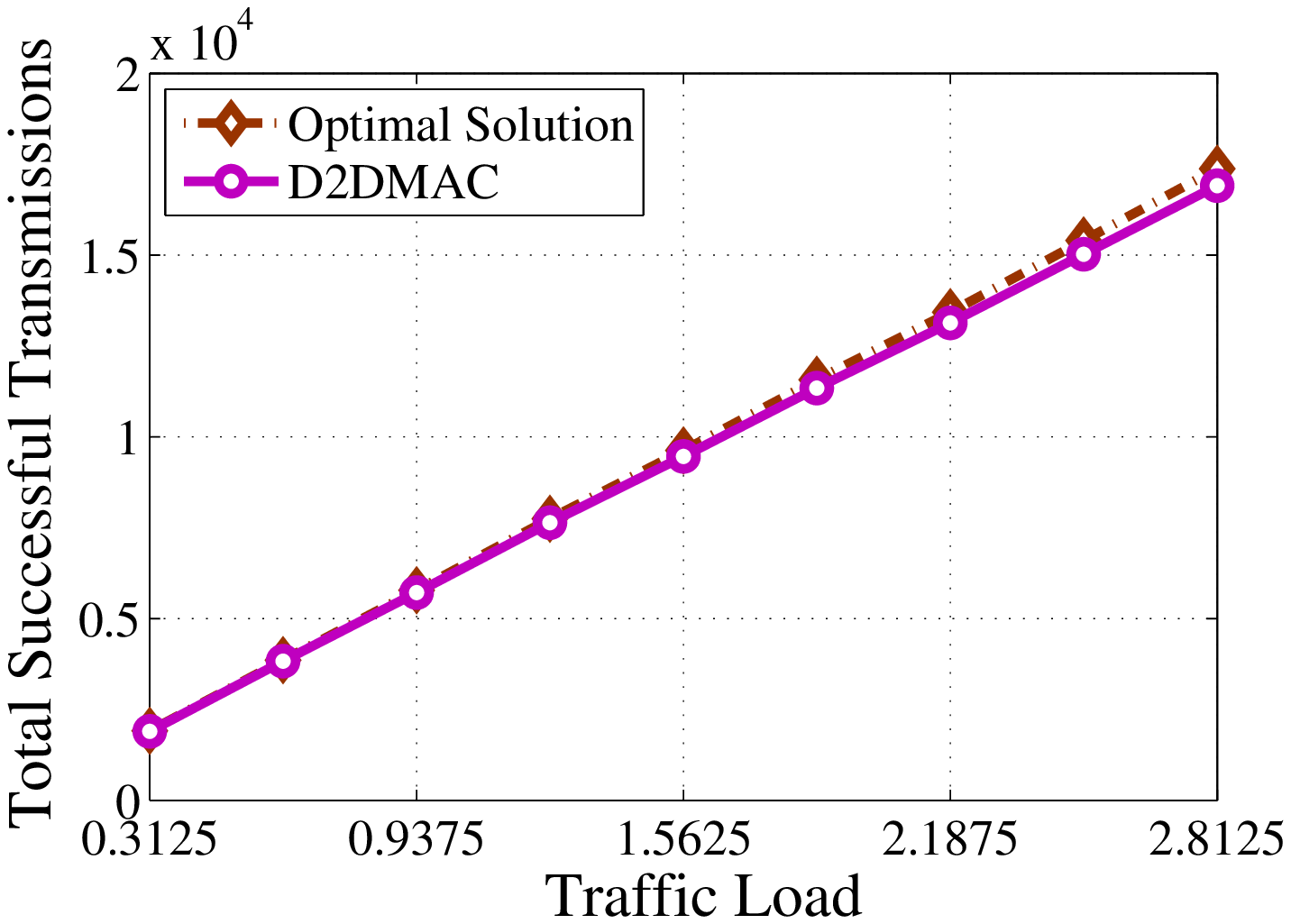}
\centerline{\small (b) Network throughput}
\end{minipage}%
\caption{Delay and throughput comparison of Optimal Solution and D2DMAC.}
\label{fig:comparison_opt} 
\vspace*{-3mm}
\end{figure}

We plot the delay and throughput comparison of D2DMAC and the Optimal Solution under Poisson
traffic in Fig. \ref{fig:comparison_opt}. From the results, we can observe that the gap between the delay of D2DMAC and Optimal Solution is negligible when the traffic load does not exceed 1.5625. With the increase of traffic load, the gap increases slowly. In
terms of network throughput, the gap between D2DMAC and the Optimal Solution is negligible. When the traffic load is 2.8125, the gap is only about 2.9\%. Therefore, we have demonstrated that D2DMAC achieves near-optimal
performance in some cases. Besides, with the path selection parameter $\beta$ optimized, the gap between D2DMAC
and Optimal Solution will decrease further.

\begin{figure}[htbp]
\begin{minipage}[t]{1\linewidth}
\centering
\includegraphics[width=0.7\columnwidth]{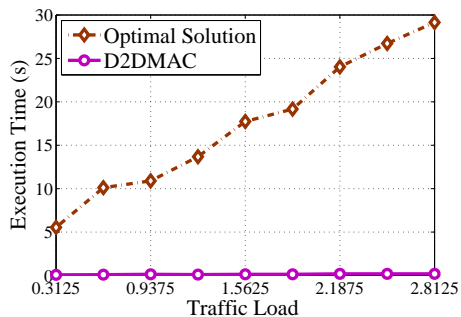}
\end{minipage}%
\caption{Execution time comparison of Optimal Solution and D2DMAC.}
\label{fig:time_opt} 
\vspace*{-3mm}
\end{figure}

We also plot the average execution time of D2DMAC and Optimal Solution under Poisson traffic in
Fig. \ref{fig:time_opt}. We can observe that
Optimal Solution takes much longer execution time than D2DMAC, and the gap increases with the
traffic load, which indicates D2DMAC has much lower computational complexity.

\subsection{Comparison with Other Protocols}

\begin{figure}[htbp]
\begin{minipage}[t]{0.5\linewidth}
\centering
\includegraphics[width=1\columnwidth,height=1.3in]{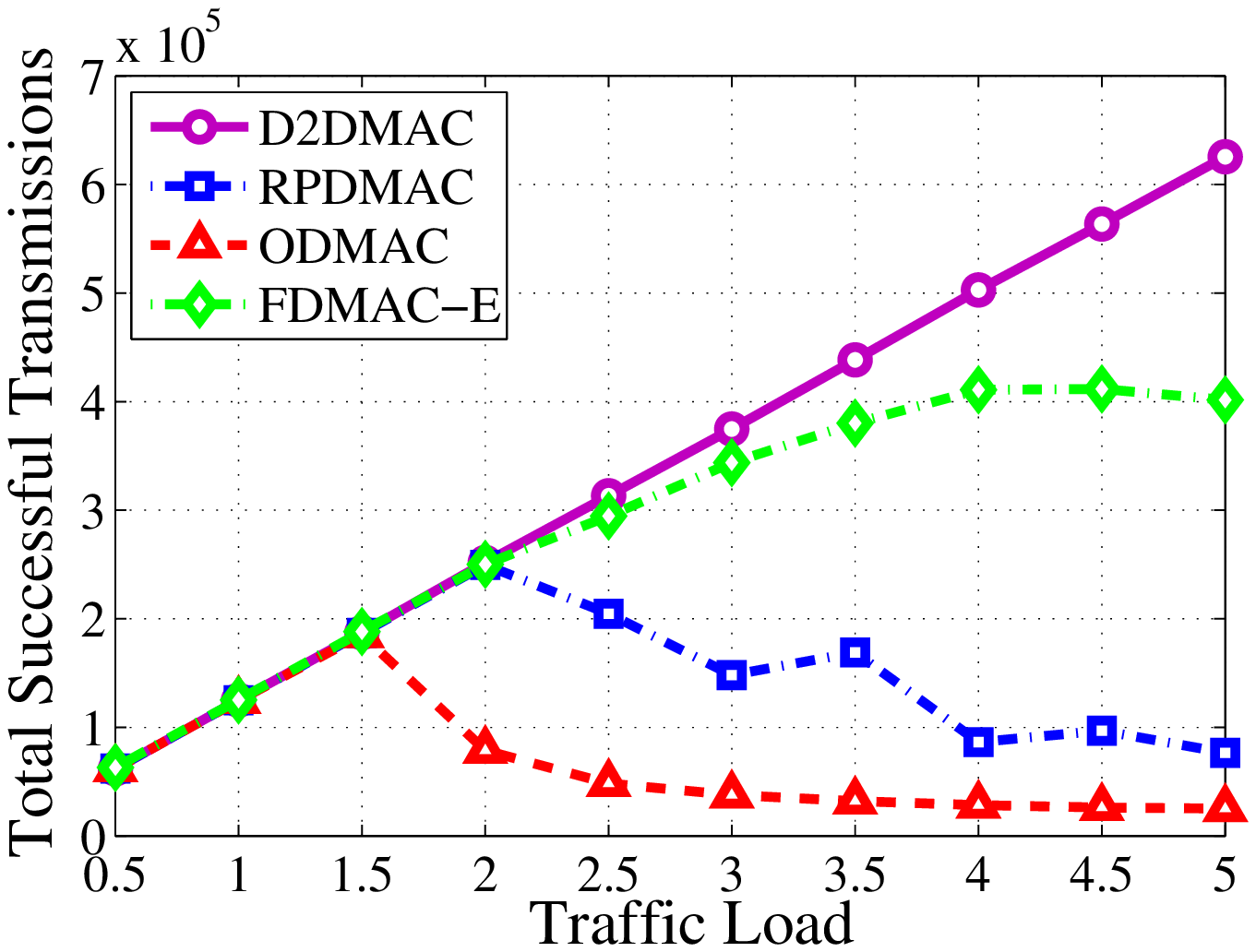}
\centerline{\small (a) Poisson traffic}
\end{minipage}%
\begin{minipage}[t]{0.5\linewidth}
\centering
\includegraphics[width=1\columnwidth,height=1.3in]{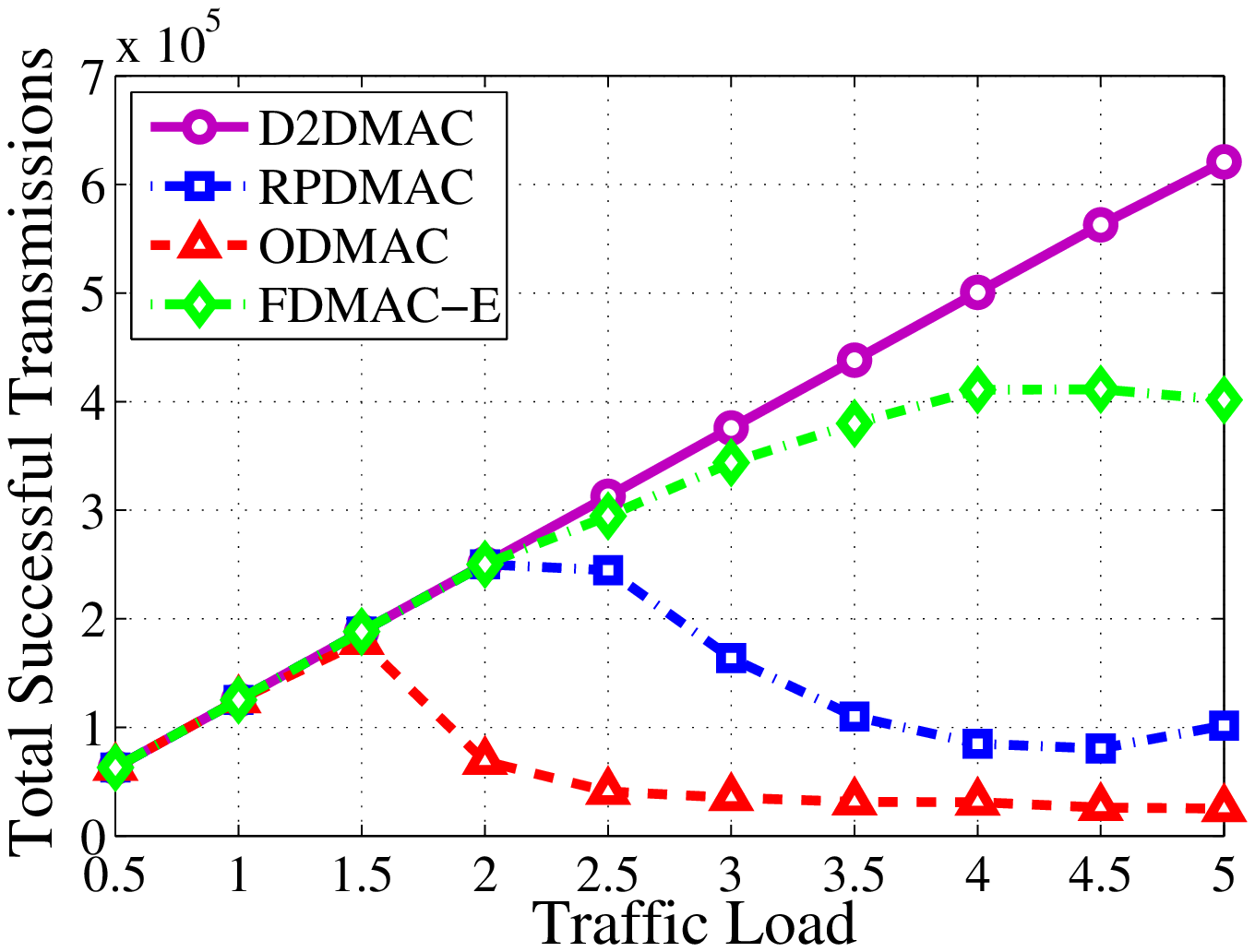}
\centerline{\small (b) IPP traffic}
\end{minipage}%
\caption{Network throughput of four protocols under Poisson and IPP traffic.}
\label{fig:comparison} 
\vspace*{-3mm}
\end{figure}

We plot the network throughput of D2DMAC, RPDMAC, ODMAC, and FDMAC-E under different traffic loads in Fig.
\ref{fig:comparison}, where the path selection parameter $\beta$ is also set to 2. As we can
observe, under light load from 0.5 to 1.5, the throughput of the four protocols is almost the same.
However, in terms of the throughput, ODMAC starts to drop at the traffic load of 1.5, RPDMAC starts to drop at the traffic load of 2, and FDMAC-E starts to drop at the traffic load of 4.5. Conversely, the throughput of D2DMAC increases
linearly with the traffic load. Under Poisson traffic, D2DMAC outperforms RPDMAC by more than
7x at the traffic load of 5. Since direct transmissions between devices are not enabled in
ODMAC, the gap between D2DMAC and ODMAC is even larger than that between D2DMAC and RPDMAC, which
indicates device-to-device transmissions can improve network throughput significantly. The throughput of D2DMAC and FDMAC-E starts to diverge at the traffic load of 2.5, and the gap increases with the traffic load. When the traffic load is 5, D2DMAC increases the network throughput by about 55.8\% compared with FDMAC-E under Poisson traffic. Since the transmission paths of FDMAC-E are the same as those of D2DMAC, the gap is caused by the joint scheduling of access and backhaul links in D2DMAC. Therefore, it also demonstrates the great benefits of the joint scheduling of the access and backhaul
networks in terms of enhancing the system performance.

\begin{figure}[htbp]
\begin{minipage}[t]{0.5\linewidth}
\centering
\includegraphics[width=1\columnwidth,height=1.3in]{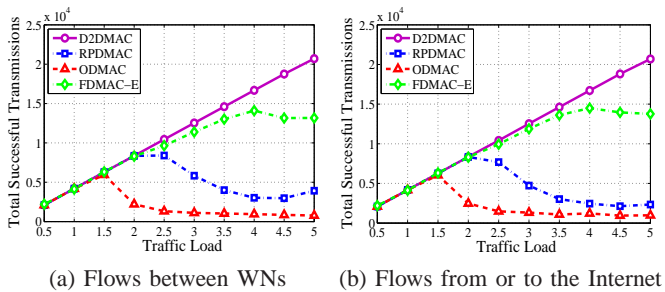}
\centerline{\small (a) Flows between WNs}
\end{minipage}%
\begin{minipage}[t]{0.5\linewidth}
\centering
\includegraphics[width=1\columnwidth,height=1.3in]{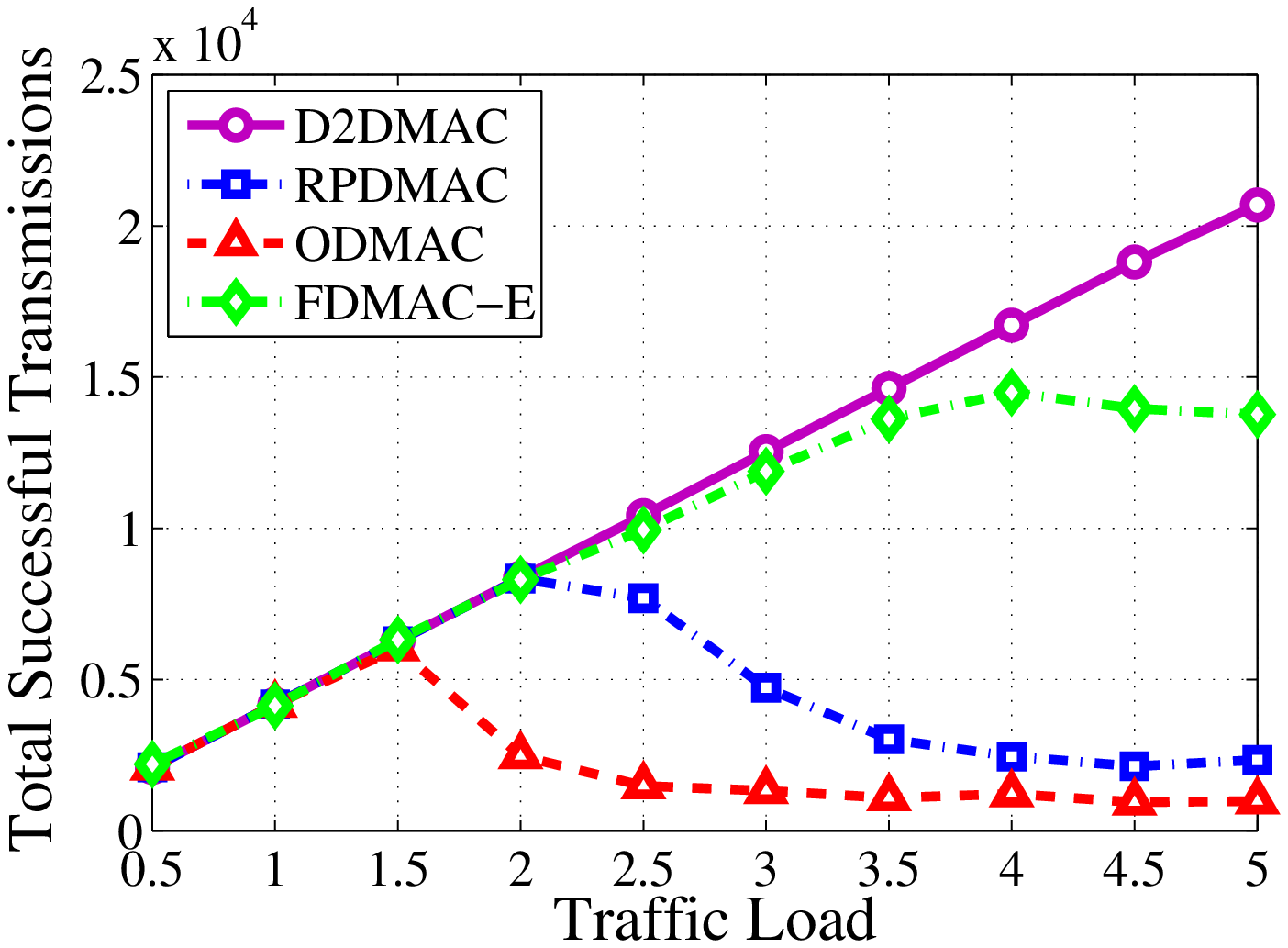}
\centerline{\small (b) Flows from or to the Internet}
\end{minipage}%
\caption{Average flow throughput of four protocols under IPP traffic.}
\label{fig:comparison_flow} 
\vspace*{-3mm}
\end{figure}

We further plot the average flow throughput of four protocols for IPP traffic under different traffic
loads in Fig. \ref{fig:comparison_flow}. These results are consistent with those in Fig.
\ref{fig:comparison}, and indicate D2DMAC improves the flow throughput of flows both between
devices and from or to the Internet significantly compared with RPDMAC, ODMAC, and FDMAC-E.

\begin{figure}[htbp]
\begin{minipage}[t]{0.5\linewidth}
\centering
\includegraphics[width=1\columnwidth,height=1.3in]{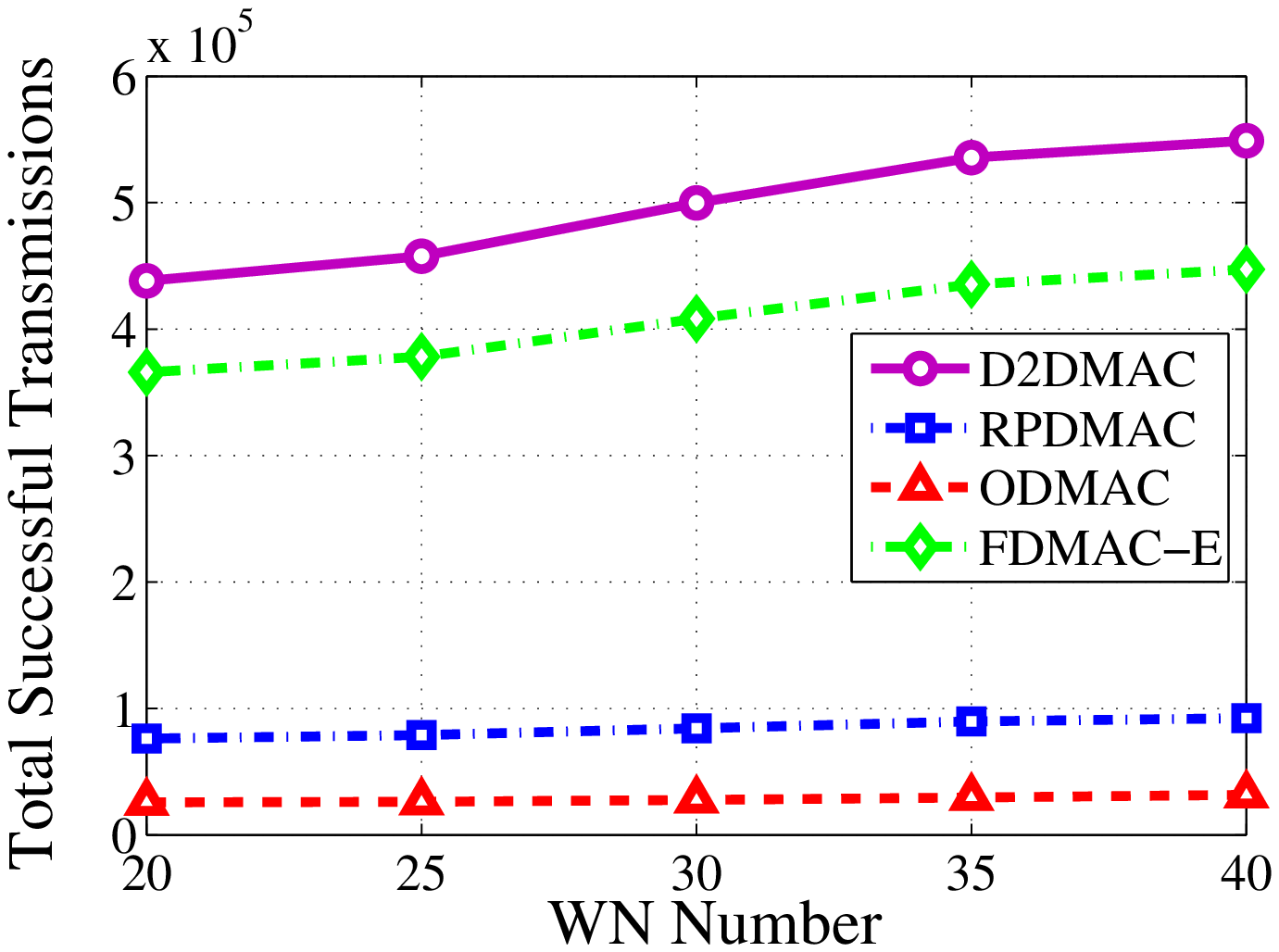}
\centerline{\small (a) Poisson traffic}
\end{minipage}%
\begin{minipage}[t]{0.5\linewidth}
\centering
\includegraphics[width=1\columnwidth,height=1.3in]{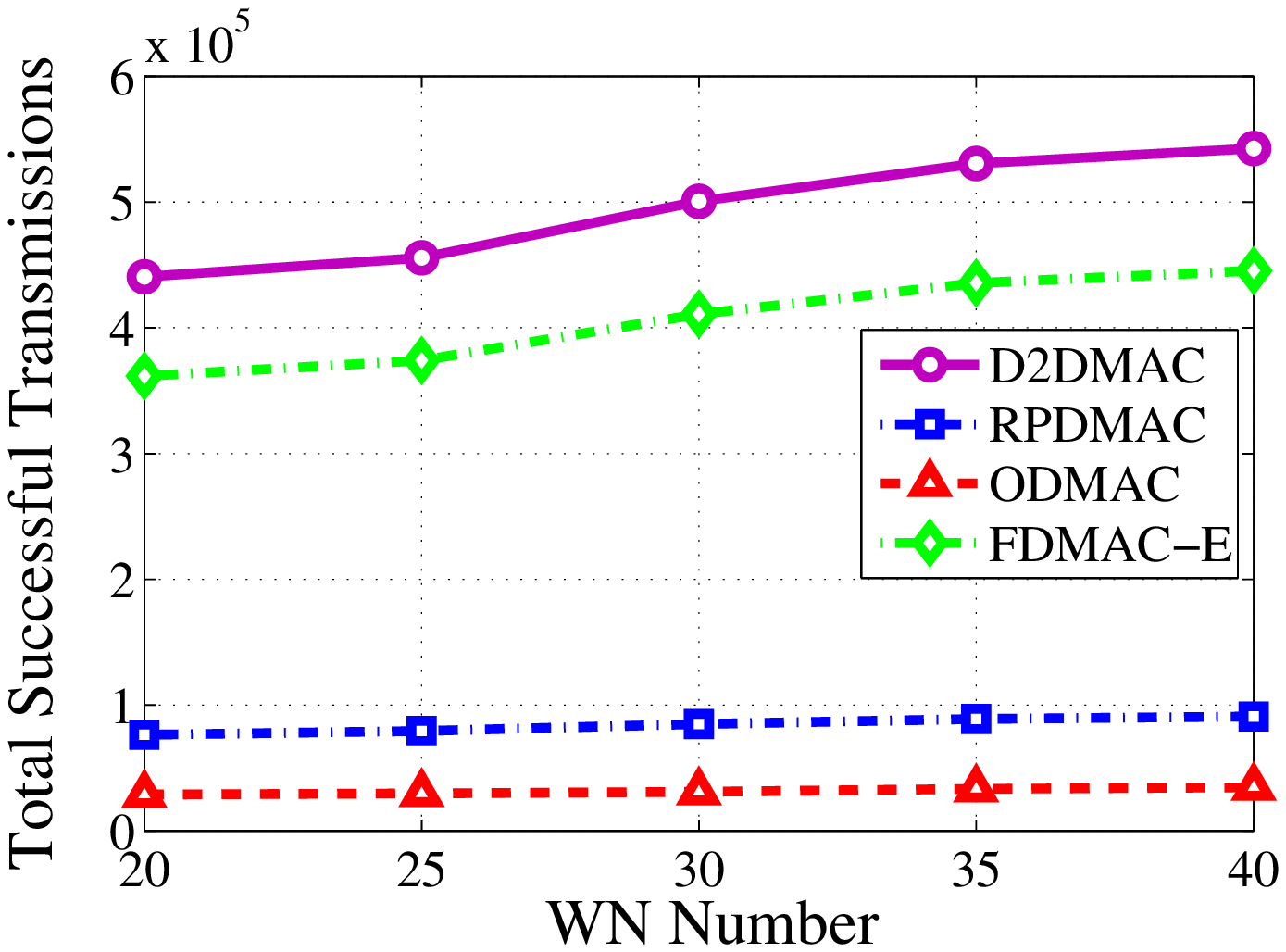}
\centerline{\small (b) IPP traffic}
\end{minipage}%
\caption{Network throughput of four protocols under different number of WNs.}
\label{fig:comparison_usernum} 
\vspace*{-3mm}
\end{figure}

To analyze the impact of user (WN) distribution density on the performance of D2DMAC, we investigate five cases of user deployment, i.e., 20, 25, 30, 35, and 40 WNs uniformly distributed in a square area of 50 m $ \times$
50 m. The traffic load is set to 4. In Fig. \ref{fig:comparison_usernum}, we plot the network throughput of four protocols with different number of WNs. From the results, we can observe that D2DMAC outperforms the other three protocols significantly. The network throughput of D2DMAC increases with the number of WNs. The reason is that with the increase of WN, the WN distribution density increases, and the average distance between nodes decreases. In this case, since the backhaul network does not change, there are more flows transmitted through the direct transmission path. Furthermore, the transmission rates of the direct transmission paths also increase due to shorter link length. Thus, the network throughput of D2DMAC increases when the number of WNs increases. As stated before, the gap between D2DMAC and FDMAC-E also indicates the advantages of joint scheduling of access and backhaul networks in D2DMAC.

\begin{figure}[htbp]
\begin{minipage}[t]{0.5\linewidth}
\centering
\includegraphics[width=1\columnwidth,height=1.3in]{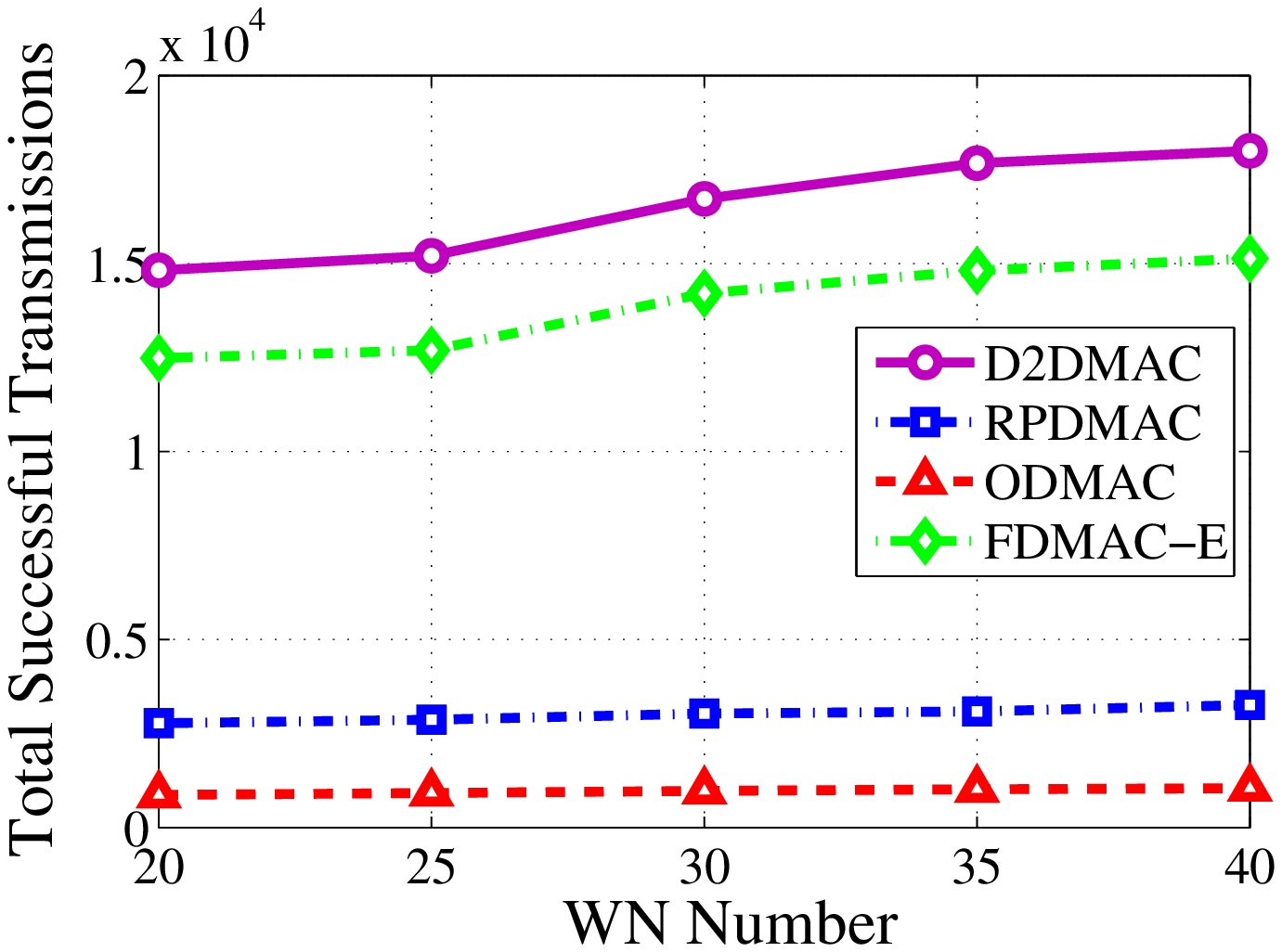}
\centerline{\small (a) Flows between WNs}
\end{minipage}%
\begin{minipage}[t]{0.5\linewidth}
\centering
\includegraphics[width=1\columnwidth,height=1.3in]{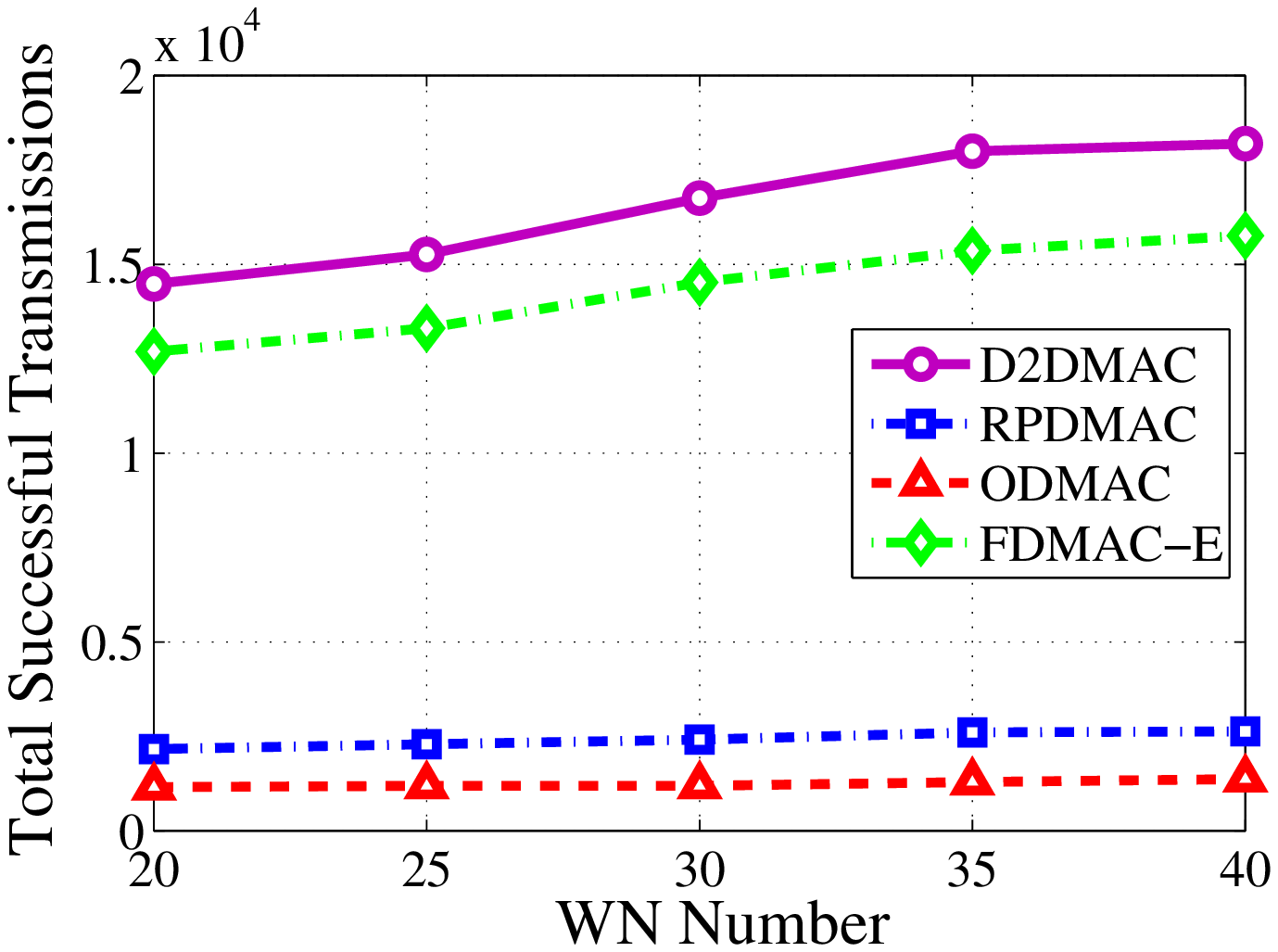}
\centerline{\small (b) Flows from or to the Internet}
\end{minipage}%
\caption{Average flow throughput of four protocols under IPP traffic.}
\label{fig:comparison_flow_usernum} 
\vspace*{-3mm}
\end{figure}

In Fig. \ref{fig:comparison_flow_usernum}, we also plot the average flow throughput of four protocols under different number of WNs. From the results, we can observe that they are consistent with those in Fig. \ref{fig:comparison_usernum}, which demonstrates D2DMAC has better utilization of the device-to-device transmissions to improve network performance under the user densely distributed scenario. Combining the results of Fig. \ref{fig:infer_radii_2}
(a), we have demonstrated that D2DMAC has superior performance in the user densely distributed scenario of the next generation mobile broadband.

\subsection{Performance under Different Path Selection Parameters } \label{S6-2}

We evaluate the performance of D2DMAC under different path selection parameters. We investigate
four cases, with $\beta$ equal to 1, 2, 3, 4, and 5 respectively. For simplicity, we denote these
cases by D2DMAC-1, D2DMAC-2, D2DMAC-3, D2DMAC-4, and D2DMAC-5.

\subsubsection{Delay}

\begin{figure}[htbp]
\begin{minipage}[t]{0.5\linewidth}
\centering
\includegraphics[width=1\columnwidth,height=1.3in]{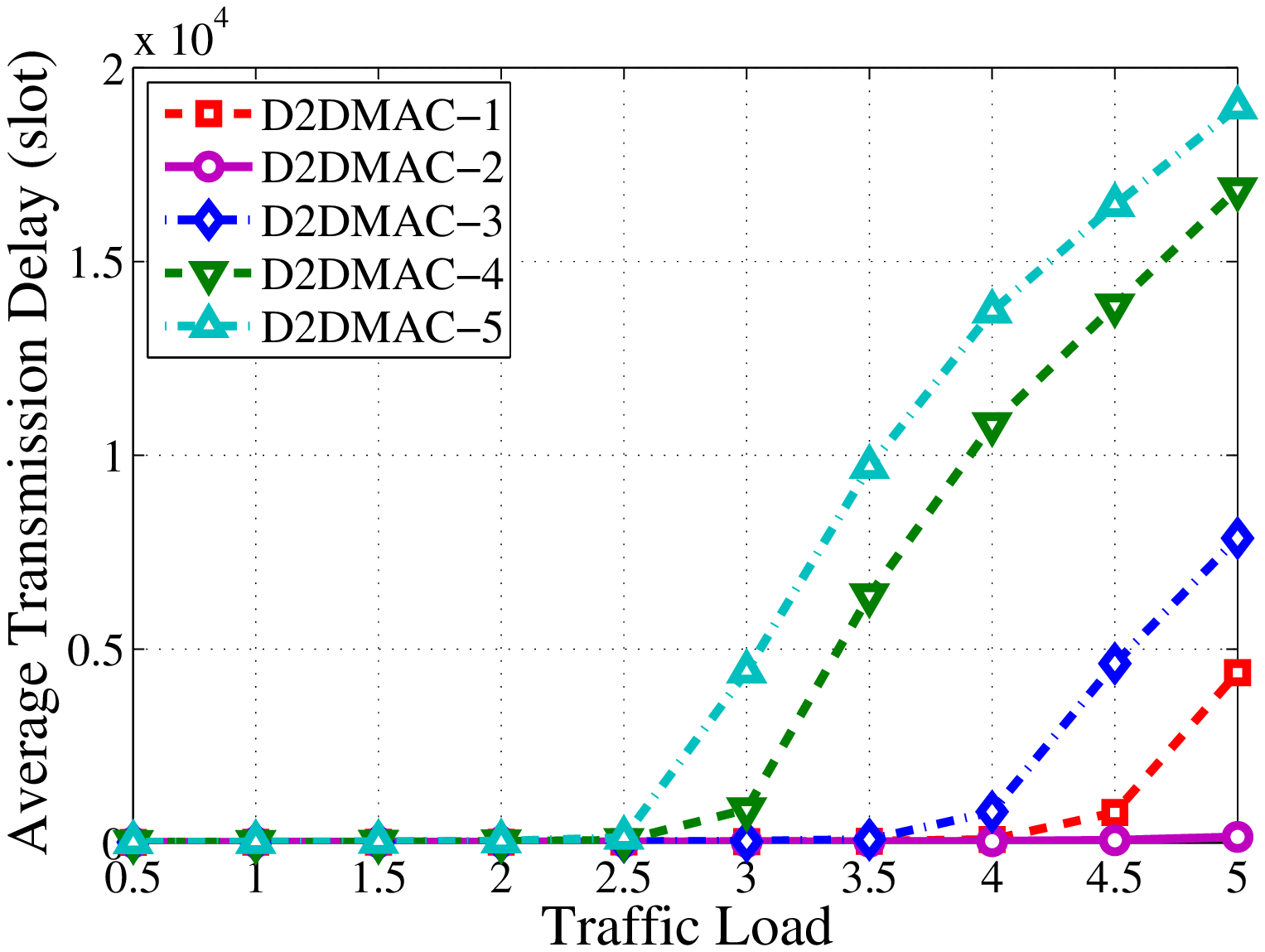}
\centerline{\small (a) Poisson traffic}
\end{minipage}%
\begin{minipage}[t]{0.5\linewidth}
\centering
\includegraphics[width=1\columnwidth,height=1.3in]{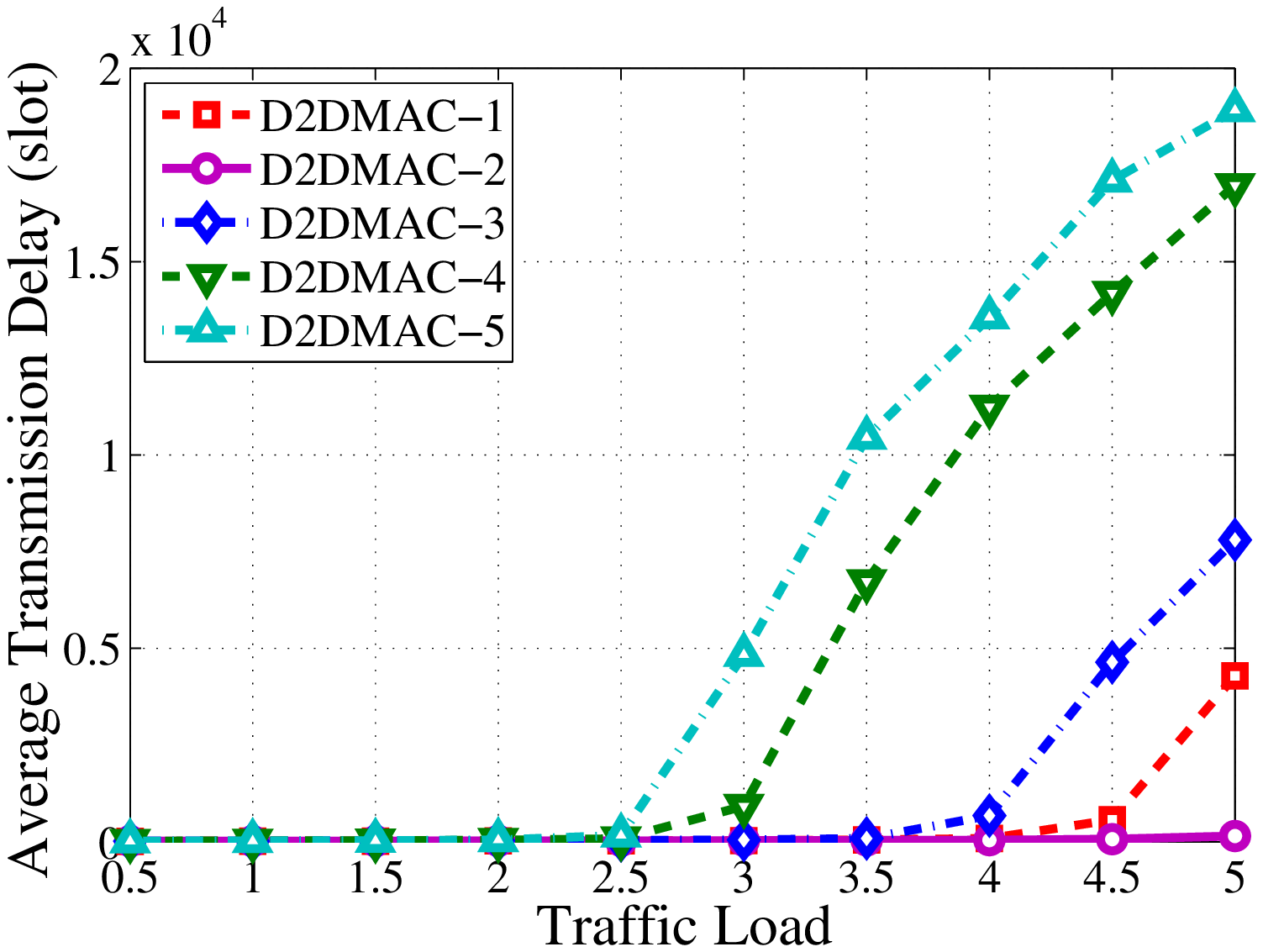}
\centerline{\small (b) IPP traffic}
\end{minipage}%
\caption{Average transmission delay of D2DMAC with different $\beta$.}
\label{fig:delay_beta} 
\vspace*{-3mm}
\end{figure}

We then evaluate the average transmission delay of D2DMAC with different path selection parameters
under different traffic loads in Fig. \ref{fig:delay_beta}. From the results, we can observe that
with the increase of traffic load, the delay of these protocols increases, and $\beta$ has a big
impact on the performance of D2DMAC. As we can observe, D2DMAC-2 achieves the best delay
performance. When $\beta$ decreases from 5 to 2, the delay becomes better. However, when $\beta$ is
equal to 1, its delay becomes worse. With $\beta$ becomes smaller, the priority of device-to-device
transmissions becomes higher. However, there is also some cases where transmission through the
backhaul network outperforms device-to-device transmission. For example, when the distance between
two devices is large, device-to-device transmission between them may be not optimal. Besides,
transmissions through the backhaul network usually have multiple hops, which may increase the
probability of concurrent transmissions for hops from different flows. Therefore, in practice,
$\beta$ should be optimized according to the actual network states and settings.

\begin{figure}[htbp]
\begin{minipage}[t]{0.5\linewidth}
\centering
\includegraphics[width=1\columnwidth,height=1.3in]{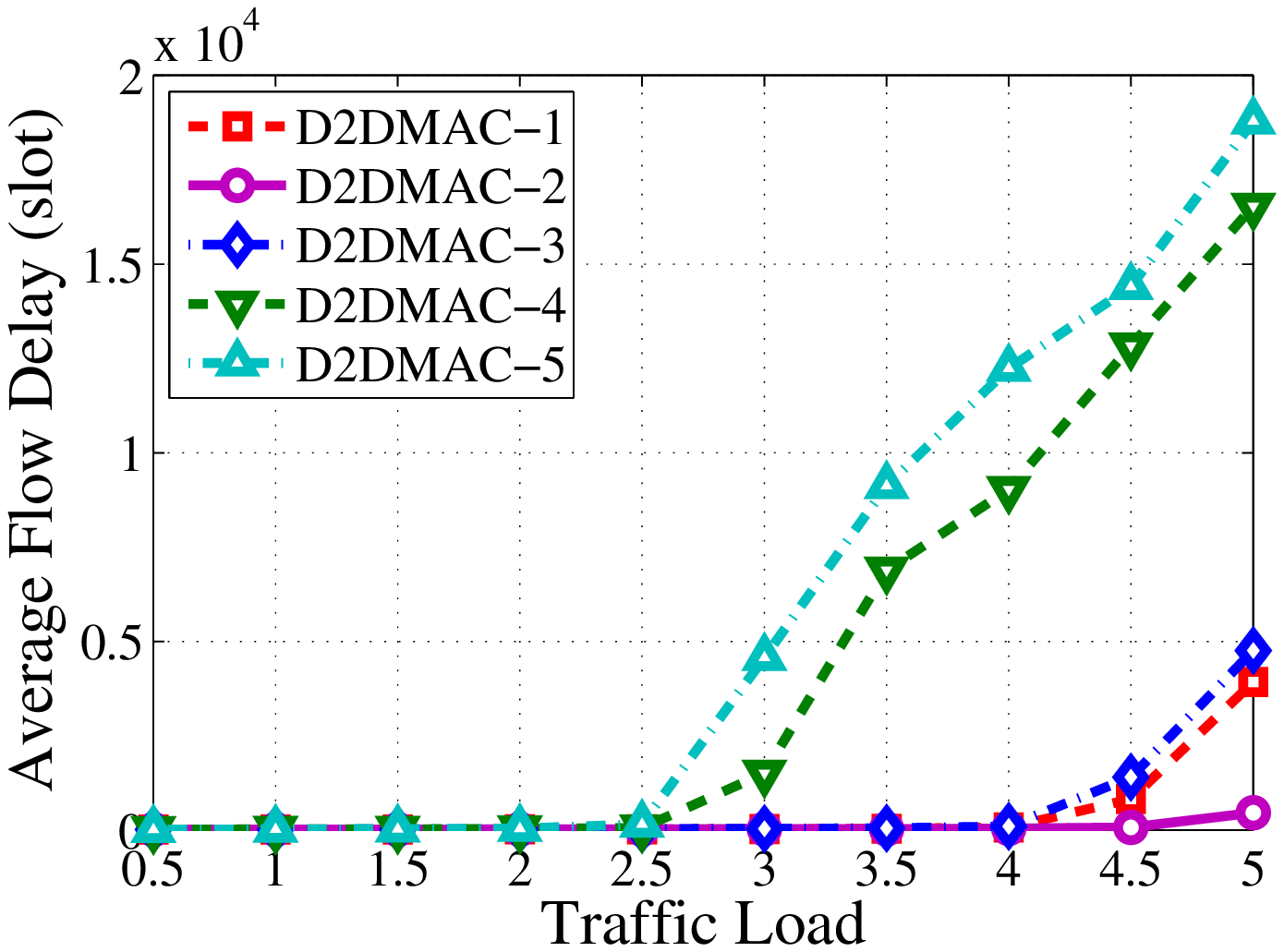}
\centerline{\small (a) Flows between WNs}
\end{minipage}%
\begin{minipage}[t]{0.5\linewidth}
\centering
\includegraphics[width=1\columnwidth,height=1.3in]{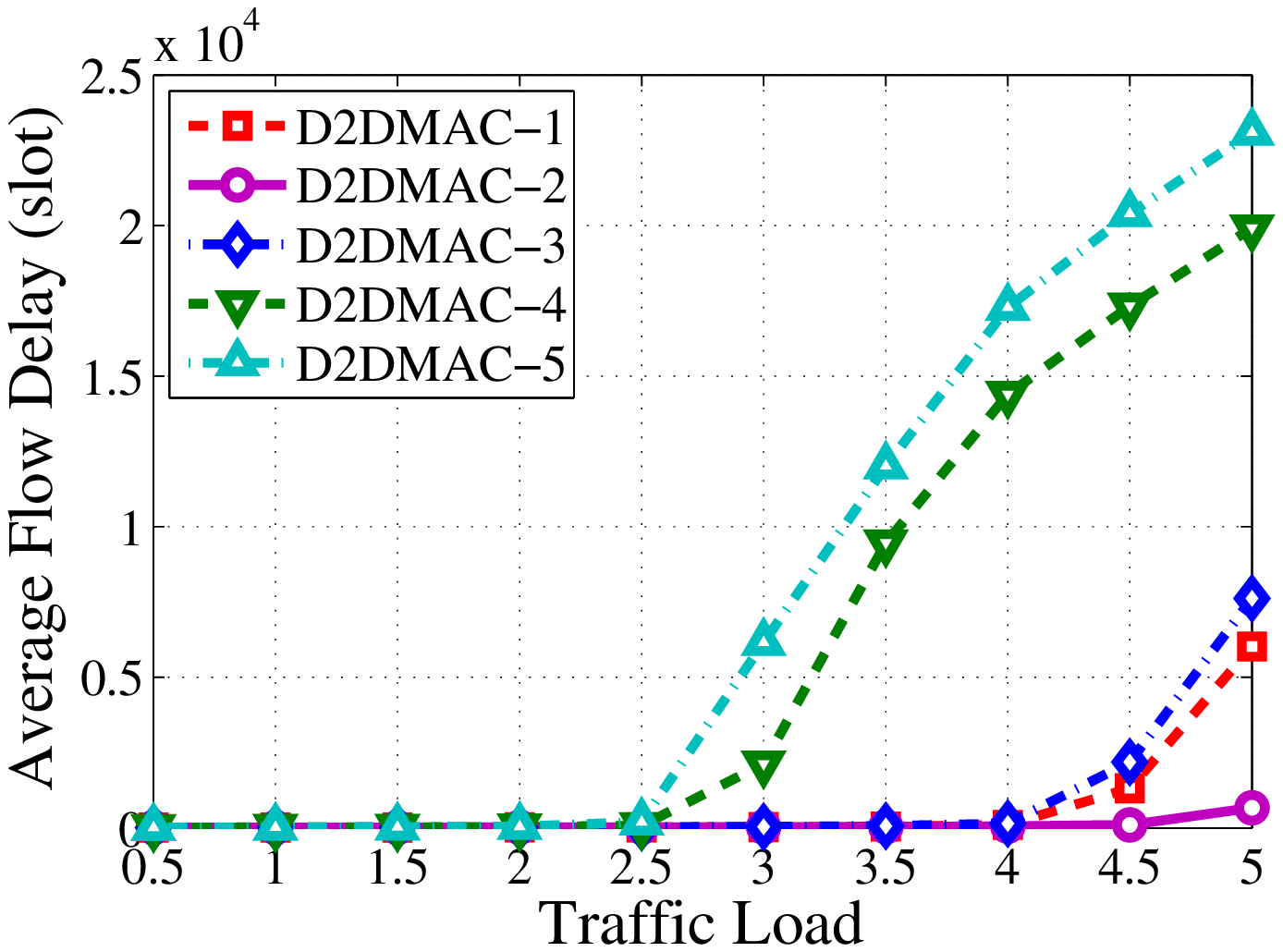}
\centerline{\small (b) Flows from or to the Internet}
\end{minipage}%
\caption{Average flow delay of D2DMAC with different $\beta$.}
\label{fig:delay_beta_ipp_flow} 
\vspace*{-3mm}
\end{figure}

We also plot average flow delay of D2DMAC with different path selection parameters under IPP
traffic in Fig. \ref{fig:delay_beta_ipp_flow}. From the results, we can observe that D2DMAC with
$\beta$ equal to 2 outperforms D2DMAC in other cases for flows both between WNs and from or to the
Internet. Therefore, the path selection parameter should be optimized to achieve low flow delay.

\subsubsection{Throughput}

\begin{figure}[htbp]
\begin{minipage}[t]{0.5\linewidth}
\centering
\includegraphics[width=1\columnwidth,height=1.3in]{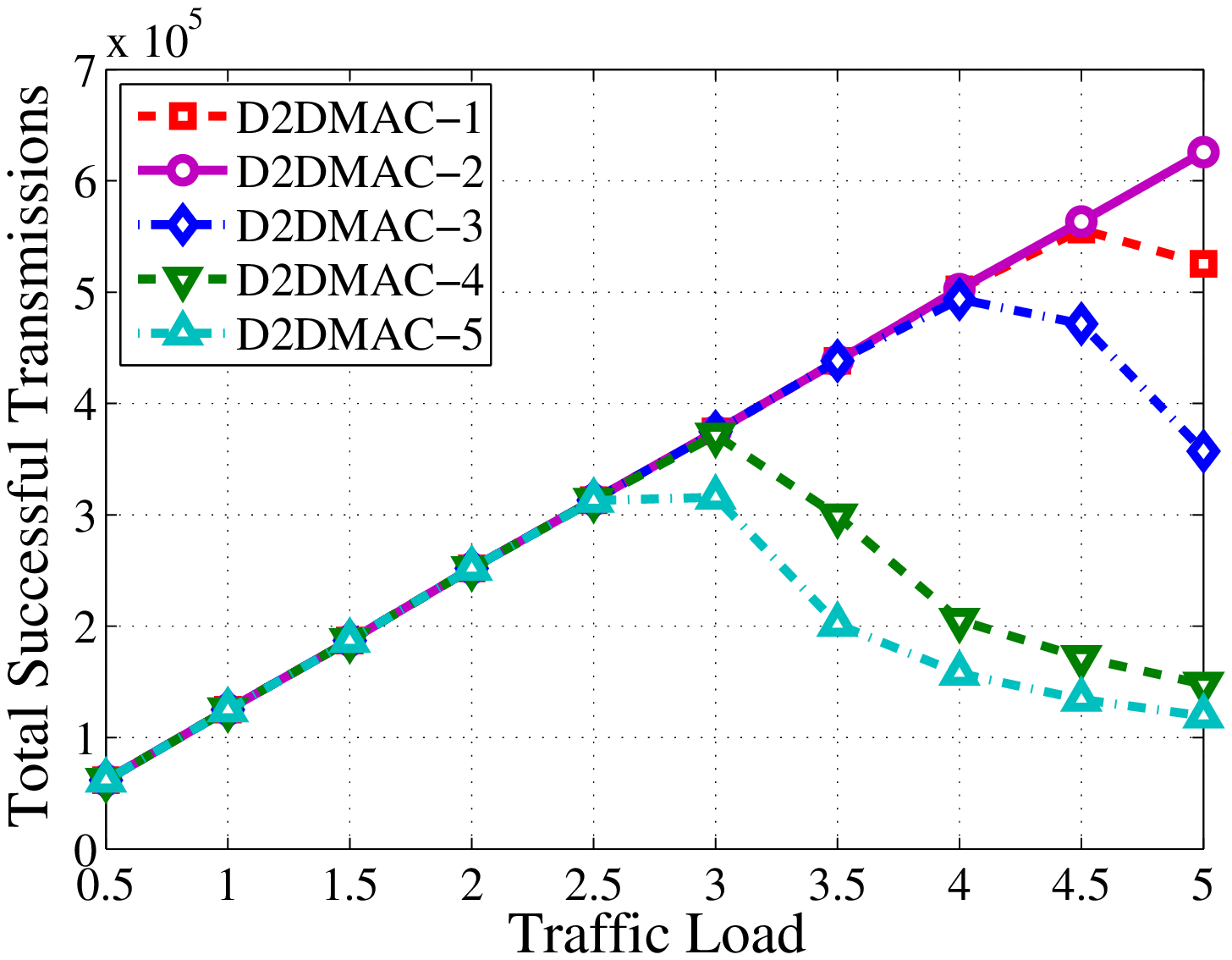}
\centerline{\small (a) Poisson traffic}
\end{minipage}%
\begin{minipage}[t]{0.5\linewidth}
\centering
\includegraphics[width=1\columnwidth,height=1.3in]{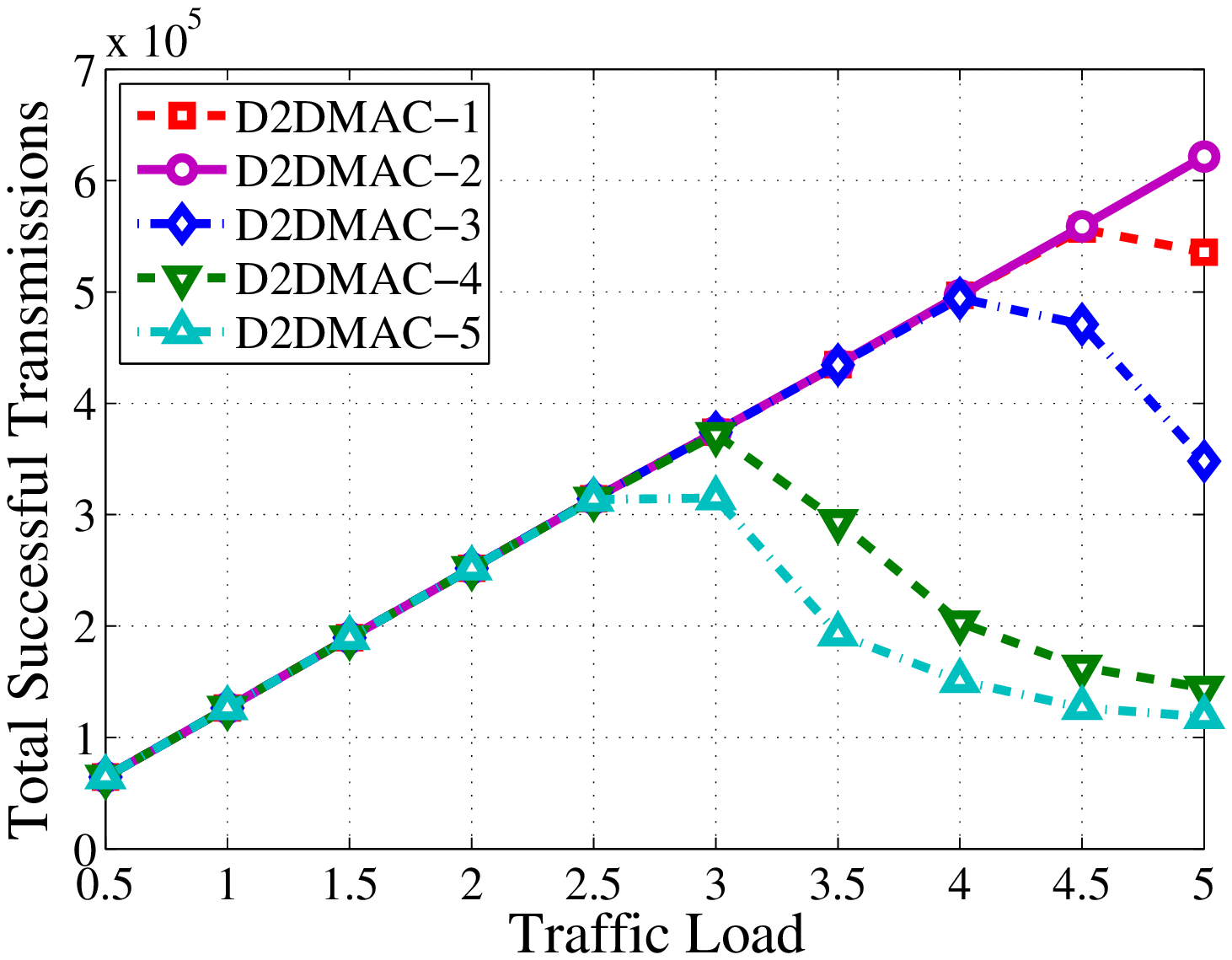}
\centerline{\small (b) IPP traffic}
\end{minipage}%
\caption{Network throughput of D2DMAC with different $\beta$.}
\label{fig:Throughput_beta} 
\vspace*{-3mm}
\end{figure}

We plot the network throughput of D2DMAC with different different path selection parameters under
different traffic loads in Fig. \ref{fig:Throughput_beta}. We can observe that the throughput is
consistent with the delay in Fig. \ref{fig:delay_beta}. Under light traffic load, throughput
increases with the traffic load. When the traffic load exceeds some critical point, which is
different for different $\beta$, the throughput starts to decrease. For example, the critical point
for D2DMAC-3 is 4. This can be explained as follows. Delay increases with the traffic load, and
when traffic load exceeds the critical point, the network becomes congested. In this case, the
delay of a considerable number of packets exceeds the threshold, and thus these packets are
discarded as failed transmissions, which leads to the abrupt drop in the throughput curve. We can
also observe that the critical point for D2DMAC-2 is larger than 5, which indicates its better
performance under heavy load.

\begin{figure}[htbp]
\begin{minipage}[t]{0.5\linewidth}
\centering
\includegraphics[width=1\columnwidth,height=1.3in]{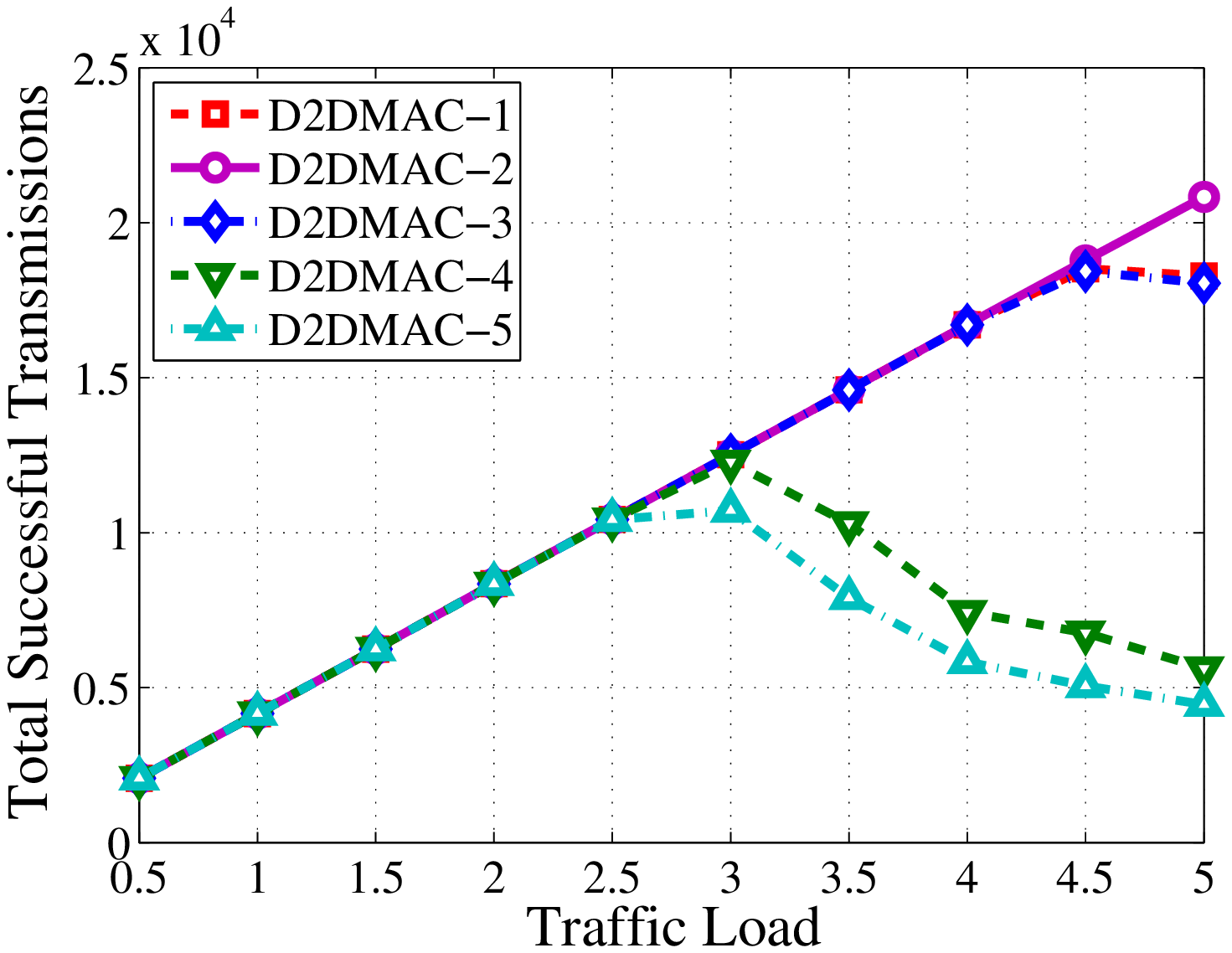}
\centerline{\small (a) Flows between WNs}
\end{minipage}%
\begin{minipage}[t]{0.5\linewidth}
\centering
\includegraphics[width=1\columnwidth,height=1.3in]{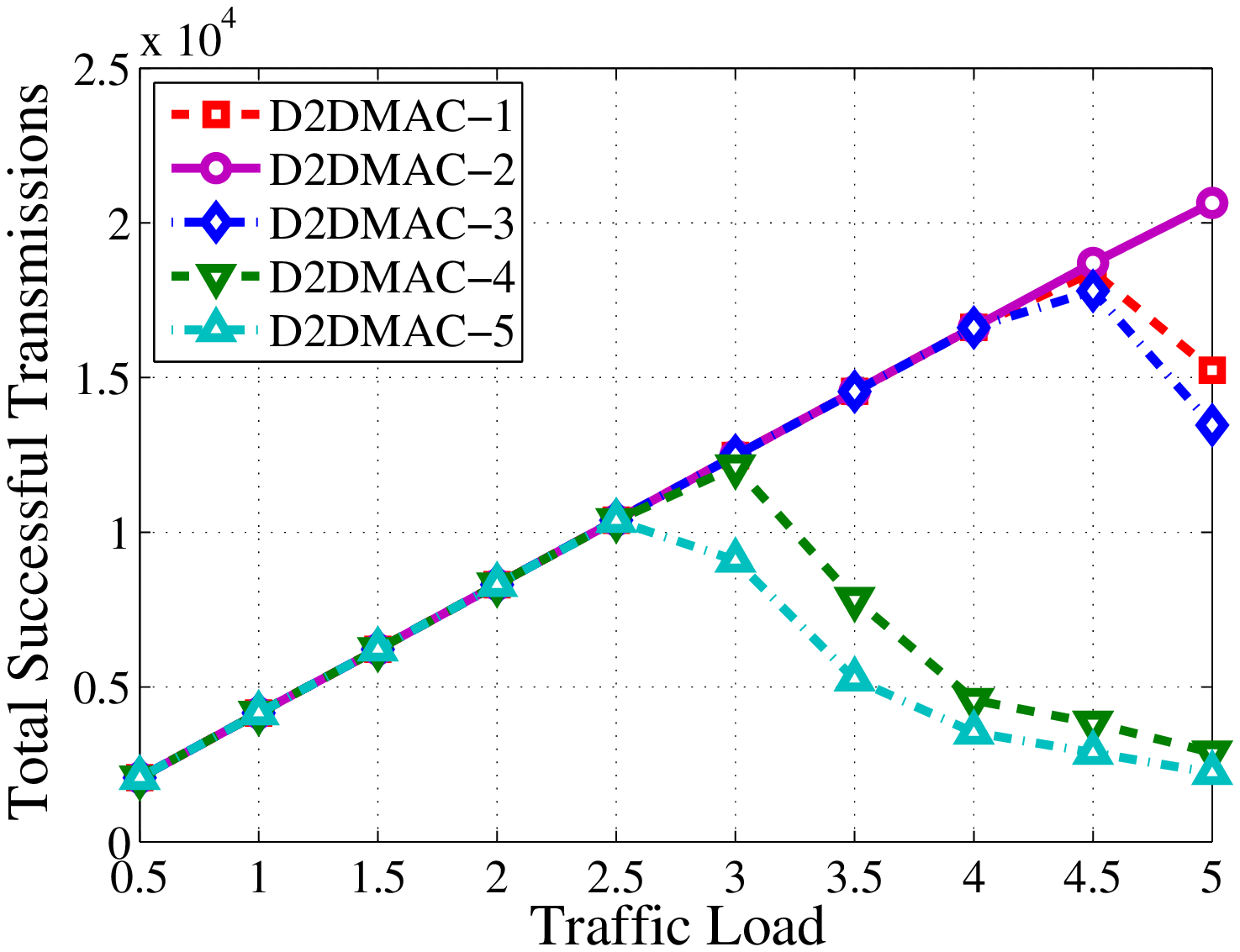}
\centerline{\small (b) Flows from or to the Internet}
\end{minipage}%
\caption{Average flow throughput of D2DMAC with different $\beta$.}
\label{fig:Throughput_beta_poisson_flow} 
\vspace*{-3mm}
\end{figure}

We also plot the average flow throughput of D2DMAC with different path selection parameters under
Poisson traffic in Fig. \ref{fig:Throughput_beta_poisson_flow}. The results indicate the importance
of choosing a suitable path selection parameter for flows both between WNs and from or to the
Internet.

In conclusion, through extensive simulations, we demonstrate D2DMAC achieves near-optimal
performance in some cases in terms of delay and throughput with low complexity, and outperforms other protocols
significantly. Furthermore, results under different path selection parameters suggest the path
selection parameter should be optimized in practice to maximize performance improvement.


\section{Conclusion}\label{S7} 

In this paper, we consider the optimal joint scheduling problem of access and backhaul networks for
small cells in the mmWave band. We propose a centralized MAC scheduling scheme, D2DMAC, where
direct transmissions between devices, i.e., device-to-device transmissions, are enabled for
performance improvement. In D2DMAC, we propose a path selection criterion to decide whether
device-to-device transmission should be enabled for each flow, and propose a transmission
scheduling algorithm to explore the potential of concurrent transmissions for maximizing spatial
reuse gain. Finally, through extensive simulations, we evaluate and analyze the delay and
throughput of D2DMAC with different path selection parameters. Comparing D2DMAC with other
protocols shows device-to-device transmissions and joint scheduling of access and backhaul networks can greatly improve network performance.
Furthermore, comparison between D2DMAC and the optimal solution demonstrates D2DMAC achieves
near-optimal performance in some cases in terms of delay and throughput. In the future work, we will develop the distributed version of D2DMAC and also consider incorporating the NLOS transmission and retransmissions into D2DMAC. Besides, we will also consider fading in the model, and investigate the impact of fading on the network performance.

\end{document}